\RequirePackage{fix-cm}
\documentclass[twocolumn,epjc3,nospthms]{svjour3}  
\smartqed  

\usepackage{physics}
\usepackage[separate-uncertainty=true, allow-number-unit-breaks=true]{siunitx}
\usepackage[version=4]{mhchem}
\usepackage{hepnames}
\usepackage{graphicx}
\usepackage{xspace}
\usepackage[dvipsnames]{xcolor}
\usepackage[colorlinks=true,citecolor=NavyBlue,filecolor=NavyBlue,linkcolor=NavyBlue,urlcolor=NavyBlue,pdftex]{hyperref}
\usepackage{cleveref}
\usepackage{amssymb} 
\usepackage[activate={true,nocompatibility},final,tracking=true,kerning=true,spacing=true,factor=1100,stretch=10,shrink=10]{microtype}
\usepackage{cite} 

\graphicspath{ {diagrams/} }
\newcommand{\echo}{ECHo\xspace}
\newcommand{\ho}{\ensuremath{\mathrm{^{163}Ho}}\xspace}
\newcommand{\dy}{\ensuremath{\mathrm{^{163}Dy}}\xspace}
\newcommand{\dystar}{\ensuremath{\mathrm{^{163}Dy^*}}\xspace}
\newcommand{\risetime}{\ensuremath{\tau_\mathrm{rise}}\xspace}

\newcommand{\mnu}{\ensuremath{m(\nu_e)}\xspace}
\newcommand{\qec}{\ensuremath{Q_\mathrm{EC}}\xspace}
\newcommand{\eec}{\ensuremath{E}\xspace}
\newcommand{\ager}{\underline{Ag}:Er\xspace}

\newcommand{\tifilter}{time information filter\xspace}
\newcommand{\psfilter}{pulse shape filter\xspace}
\newcommand{\dt}{\ensuremath{\Delta t}\xspace}
\newcommand{\dtch}{\ensuremath{\Delta t_\mathrm{ch}}\xspace}
\newcommand{\tholdoff}{\ensuremath{t_\mathrm{hold}}\xspace}
\newcommand{\tcoincidence}{\ensuremath{t_\mathrm{coinc}}\xspace}
\newcommand{\binwidth}{\ensuremath{\Delta t_\mathrm{bin}}\xspace}
\newcommand{\activitych}{\ensuremath{A_\mathrm{ch}}\xspace}
\newcommand{\nexcpect}{\ensuremath{\langle N_\mathrm{ch} \rangle}\xspace}
\newcommand{\fnoise}{\ensuremath{f_\mathrm{noise}}\xspace}

\newcommand{\figureofmerit}{\ensuremath{FOM}\xspace}
\newcommand{\pretriggernoise}{\ensuremath{\sigma}\xspace}
\newcommand{\brightness}{\ensuremath{b}\xspace}
\newcommand{\signalheight}{\ensuremath{V}\xspace}
\newcommand{\counts}{\ensuremath{N}\xspace}
\newcommand{\intensity}{\ensuremath{I}\xspace}

\newcommand{\chitworeduced}{\ensuremath{\chi^2_\mathrm{red}}\xspace}
\newcommand{\amplitude}{\ensuremath{A}\xspace}
\newcommand{\offset}{\ensuremath{O}\xspace}

\newcommand{\pit}{\ensuremath{\mathrm{PIT}}\xspace}
\newcommand{\pot}{\ensuremath{\mathrm{POT}}\xspace}

\journalname{Eur. Phys. J. C}
\begin{document}
\title{Data reduction for a calorimetrically measured \ho spectrum of the ECHo-1k experiment}

\author{Robert Hammann\thanksref{e1,addr1}
        \and
        Arnulf Barth\thanksref{e2,addr1}
        \and
        Andreas Fleischmann\thanksref{addr1}
        \and
        Dennis Schulz\thanksref{addr1}
        \and
        Loredana Gastaldo\thanksref{addr1}
}

\thankstext{e1}{e-mail: robert.hammann@kip.uni-heidelberg.de}
\thankstext{e2}{e-mail: arnulf.barth@kip.uni-heidelberg.de}

\institute{Kirchhoff-Institute for Physics, Im Neuenheimer Feld 227, 69120 Heidelberg, Germany \label{addr1}}

\date{Received: date / Accepted: date}
\maketitle
\begin{abstract}
\noindent
The electron capture in \ho experiment (ECHo) is designed to directly measure the effective electron neutrino mass by analysing the endpoint region of the \ho electron capture spectrum. We present a data reduction scheme for the analysis of high statistics data acquired with the first phase of the ECHo experiment, ECHo-1k, to reliably infer the energy of \ho events and discard triggered noise or pile-up events. On a first level, the raw data is filtered purely based on the trigger time information of the acquired signals. On a second level, the time profile of each triggered event is analysed to identify the signals corresponding to a single energy deposition in the detector. We demonstrate that events not belonging to this category are discarded with an efficiency above \SI{99.8}{\%}, with a minimal loss of \ho events of about \SI{0.7}{\%}. While the filter using the trigger time information is completely energy independent, a slight energy dependence of the filter based on the time profile is precisely characterised. This data reduction protocol will be important to minimise systematic errors in the analysis of the \ho spectrum for the determination of the effective electron neutrino mass.
\end{abstract}
\section{Introduction}
    \label{sec:Introduction}
    \begin{figure*}[!ht]
    \centering
        \includegraphics[width=0.482\textwidth]{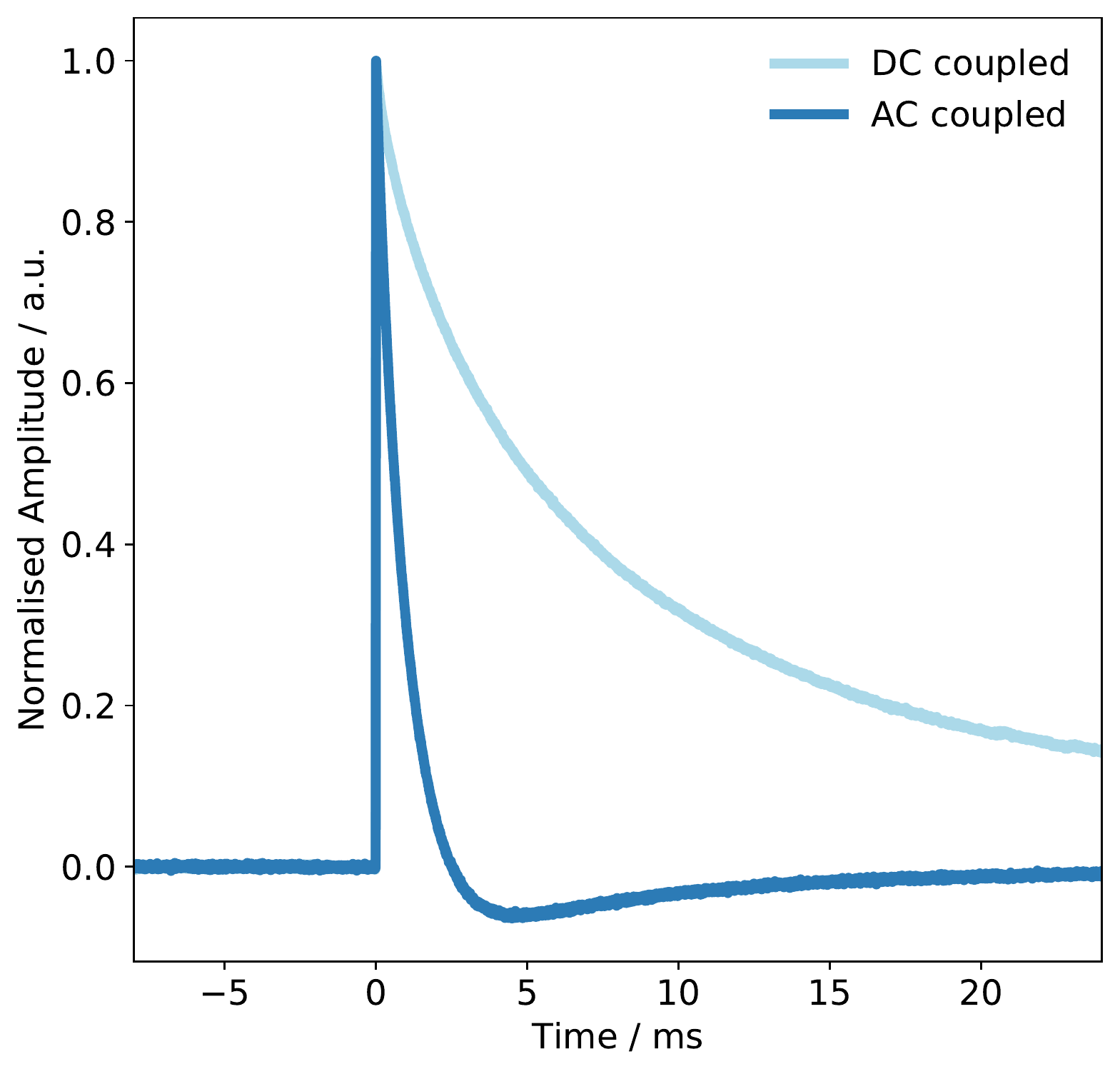}
        \includegraphics[width=0.508\textwidth]{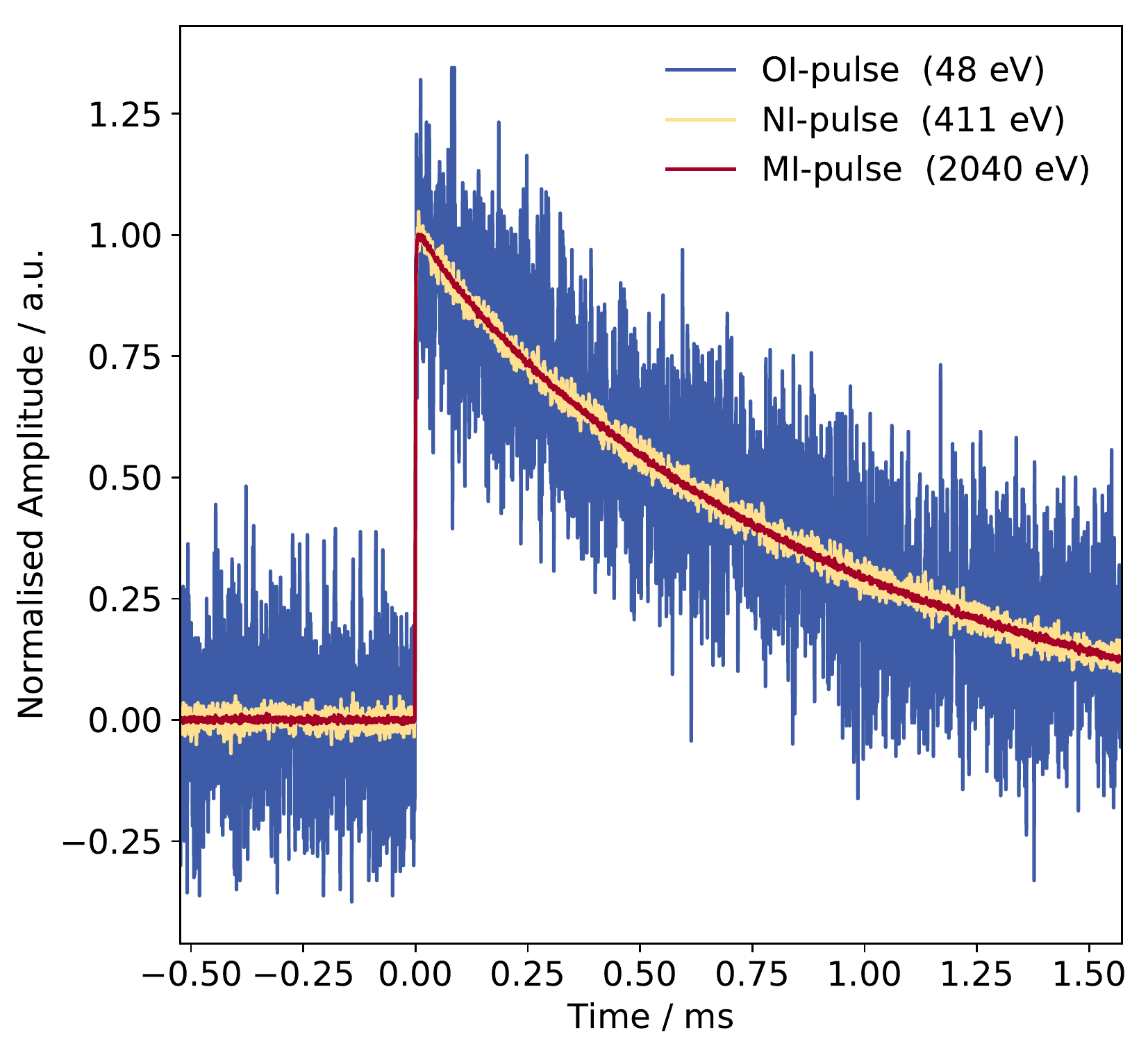}
    \caption{\protect\textit{Left:} Comparison of the pulse shape of a DC-coupled signal and an AC-coupled signal. \textit{Right:} Three time traces of different energies normalised to unity amplitude. MI, NI and OI refers to the corresponding peak in the \ho spectrum. The first quarter of the time traces are pre-trigger samples. The traces are acquired with AC-coupled signal amplification}
    \label{fig:pulse}
    \end{figure*}
    
    A promising approach for the direct determination of the effective electron neutrino mass \mnu is the analysis of the endpoint region of a calorimetrically measured spectrum following the electron capture (EC) of \ho. This method was first proposed in \cite{Rujula_82} and is presently pursued by two collaborations, \echo \cite{Gastaldo_17} and HOLMES \cite{Alpert_15}. The EC of \ho is characterised by  \qec=\SI[parse-numbers=false]{2.833 \pm 0.030 (stat) \pm 0.015 (sys)}{\kilo\electronvolt} \cite{Eliseev_15}, which is the available energy for the decay,
    and a half life of $T_{1/2}=\SI[parse-numbers=false]{4570 \pm 50}{y}$ \cite{Baisden_83}.
    
    In the EC of \ho, a shell electron is captured by the nucleus, converting a proton into a neutron and emitting an electron neutrino. The daughter atom is left in an excited state \dystar, which predominantly decays to the ground state \dy via non-radiative processes. In first approximation, the excited state \dystar is characterised by a vacancy of an inner shell electron and an additional electron in the 4f-shell. In this simplified model, the spectrum is given by six resonances centred at the binding energies of the orbitals of the captured electrons in the potential of the daughter nucleus: MI($3\mathrm{s}_{1/2}$), MII($3\mathrm{p}_{1/2}$), NI($4\mathrm{s}_{1/2}$), NII($4\mathrm{p}_{1/2}$), OI($5\mathrm{s}_{1/2}$), OII($5\mathrm{p}_{1/2}$)\footnote{Note that due to the low Q-value of this particular decay, the capture of electrons with principle quantum number $n<3$ is kinematically forbidden. Moreover, an additional PI-line cannot be observed experimentally as the $6\mathrm{s}_{1/2}$ electrons move to the electron bands of the host material due to their weak binding to the \ho atom.}. The amplitude of each resonance is defined by the phase space factor and the probability that an electron wave function in the given state overlaps with the nucleus. The phase space factor contains the information on the neutrino mass and gives rise to a cutoff at the endpoint energy $\eec = \qec - \mnu$. The endpoint region is therefore most sensitive to a finite effective electron neutrino mass. Recently, a more accurate description of the calorimetrically measured \ho spectrum has been developed. Besides the main resonances, it contains a number of structures that take into account electron scattering processes in the atom and excitations to the continuum \cite{Bra18,Bra20}.
    
    In order to obtain sub-eV sensitivity on \mnu, one must measure energies below \SI{3}{\kilo\electronvolt} with eV-precision. In addition, a good intrinsic time resolution $\sim$\,\SI{100}{\nano\second} and a high statistics of more than $10^{14}$ \ho events are essential \cite{Gastaldo_17}. For \echo, metallic-magnetic calorimeters (MMCs) operated at temperatures below \SI{30}{\milli\kelvin} \cite{Fle05} inside a dry dilution refrigerator\footnote{Produced by BlueFors Cryogenics Oy, Arinatie 10, 00370 Helsinki, Finland.} are used to meet these requirements. The particular type of MMC used for \echo is characterised by a particle absorber that encloses the high-purity \ho source. If an energy \eec is deposited in the absorber, its temperature rises with a time constant $\risetime\sim\SI{100}{\nano\second}$. A paramagnetic \ager temperature sensor, which is situated in an external static magnetic field and is thermally well coupled to the absorber, acts as a precise thermometer. The magnetisation of this sensor is temperature dependent. Consequently, a change in temperature causes a change of magnetic flux in a suitable pick-up coil. A flux-locked-loop dc-SQUID (direct current - superconducting quantum interference device) readout is then used to convert the change of flux into a change of voltage proportional to the initially deposited energy \eec. A gold thermal link made of several gold films with increasing width finally connects the detector to an on-chip thermal bath so that the initial temperature is restored. At the operating temperature of \SI{20}{\milli\kelvin}, the recovery time is of the order of milliseconds. The decaying part of the temperature pulse can be described by a sum of exponential functions due to the step structure of the thermal link to the on-chip thermal bath. The rising part of the pulse can be affected by a reduced readout bandwidth, which effectively increases the signal rise time \risetime. For a DC-coupled signal, the time constants with their respective amplitudes fully specify the shape of a thermal pulse. AC coupling of the signal keeps the baseline offset at \SI{0}{\volt}. This strongly modifies the signal shape as shown in \cref{fig:pulse} (left).
    
    The detector geometry used for \echo is a double meander, which corresponds to two superconducting meander structures connected in parallel with the input coil of one dc-SQUID. On top of each meander, a paramagnetic sensor is fabricated. To polarise the spins in the sensor, a constant magnetic field is generated by a persistent current in the meander structures. Simultaneously, the meander structure serves as a readout coil to detect the magnetisation changes in the sensor. In such a gradiometric setup, the signal of a common change in temperature in the two sensors cancels out, which significantly reduces noise caused by global temperature fluctuations of the chip. On top of each sensor, a gold absorber with the dimensions \SI{180}{\micro\meter} x \SI{180}{\micro\meter} x \SI{10}{\micro\meter} is fabricated. Each set comprising meander, sensor and absorber is referred to as one pixel and one gradiometer consisting of two pixels is referred to as one detector, which is read out by a two-stage SQUID setup \cite{Drung07}. Thus, each detector is associated to one readout channel (detector channel). Due to the opposite polarity of the screening current in the double meander structure of the detectors, the voltage signals from the two pixels of one detector have opposite signs. The triggered signals are separated into positive and negative polarity pulses based on the voltage slope after the trigger. Thus, signals from the two pixels of one detector can be distinguished.
    
    In the ECHo-1k high statistics measurement (run 24 and run 25), two ECHo-1k chips \cite{Mantegazzini2021_phd,Mantegazzini21_unpub} have been used. Each chip hosts 32 detectors implanted with \ho, two double meanders to study the properties of non-implanted pixels and two so-called temperature channels, which feature only one sensor and are therefore sensitive to temperature fluctuation of the substrate. In 7 of the 32 implanted detectors, only one pixel contains \ho to allow for in-situ background measurements. Signals from a total of 68 pixels have been acquired over a period of five months, 58 of which are implanted with \ho. The average activity per detector is approximately \SI{1}{Bq}. The data reduction scheme discussed in this work has been developed to eliminate spurious events like triggered noise or pileup from these ECHo-1k datasets.
    
    The signals of each detector channel are amplified by a room temperature SQUID electronics\footnote{SQUID electronics type XXF-1 from Magnicon GmbH, Hamburg}, controlling the two-stage SQUID readout, and digitised by a 16-channel analogue-to-digital converter (ADC)  with 16-bit resolution and a maximum sampling rate of \SI{125}{\mega\hertz}\footnote{SIS3316 from Struck Innovative Systeme} \cite{Mantegazzini21}. To generate a trigger, a trapezoidal finite impulse response (FIR) filter is employed. The trigger threshold can be chosen individually for each detector channel and is usually set to be just above perceived noise levels. Once a signal is triggered, a time trace of 16384 voltage samples is saved, the first quarter of which is dedicated to pre-trigger samples. Three examples of saved AC-coupled \ho traces are shown in \cref{fig:pulse} (right). For a sampling rate of \SI{125}{\mega\hertz} and an oversampling of 16, i.e. a difference between two samples of $16/\SI{125}{MHz} = \SI{128}{ns}$, the total saved time window of each trace is \SI{2.09715}{\milli\second}. During this time period after a trigger, no further triggers from the same detector channel are accepted. For each trace, the timestamp of its trigger is saved. One can then calculate the time difference \dt to the previously saved trace in any of the acquiring detector channels of one ECHo-1k chip, and the time difference to the previously saved trace within the same detector channel, \dtch.
    
    The information extracted from the analysis of the timestamps is used for reliable and energy-independent data reduction. This is crucial to avoid distortions of the spectrum, particularly in the endpoint region. The shapes of the filtered traces are then analysed to remove remaining spurious signals. The method used is based on the chi-squared goodness of fit measure, which is calculated for each event following a template fit.
    
    In the first part, we present different signal families that have been identified in our data. In the second part, methods to eliminate various spurious events are discussed and the performance of these algorithms applied to a subset of ECHo-1k data is evaluated in the last part.

\section{Signal Families}
\label{sec:signalFamilies}
\begin{figure*}[!ht]
\centering
    \includegraphics[width=\textwidth]{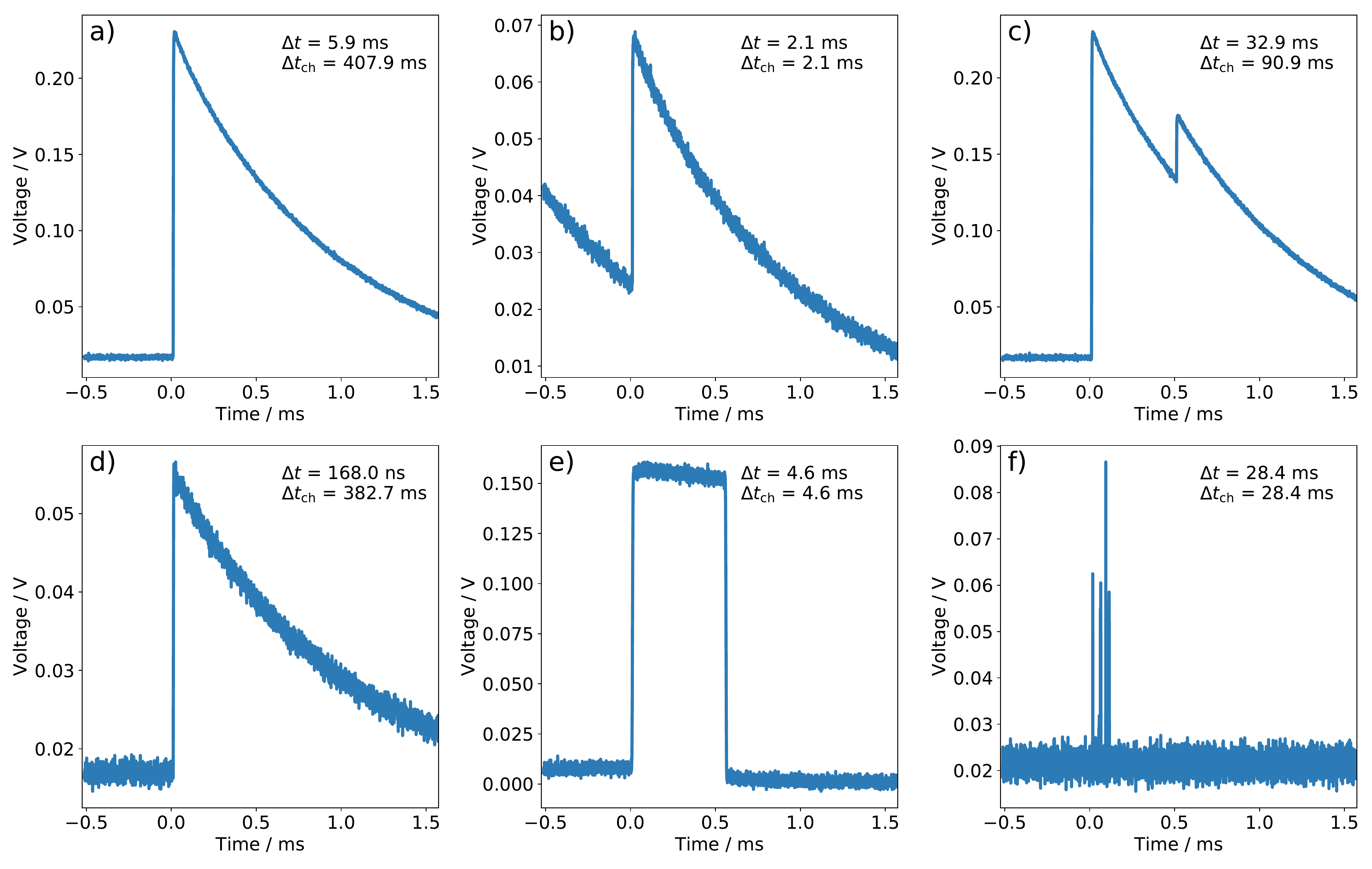}
\caption{a) \ho event b) Pileup-on-tail outside the time window (\pot) c) Pileup-on-tail with both signals inside the time window (\pit) d) Pixel-pixel coincidence event with thermal pulse shape e) GSM Signal f) Triggered noise}
\label{fig:pulseshapes}
\end{figure*}

\begin{figure*}[!ht]
    \centering
    \includegraphics[width=0.49\textwidth]{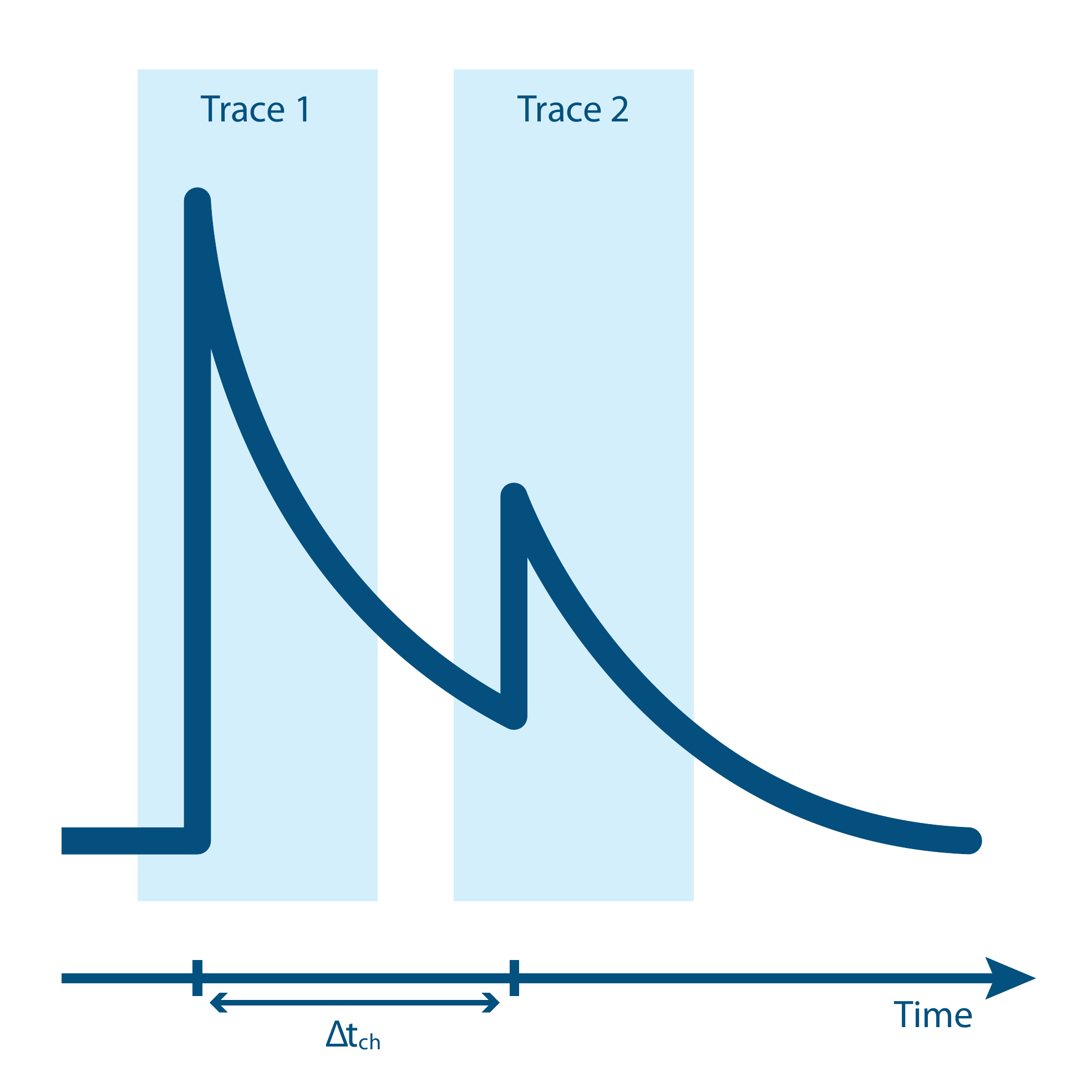}
    \includegraphics[width=0.49\textwidth]{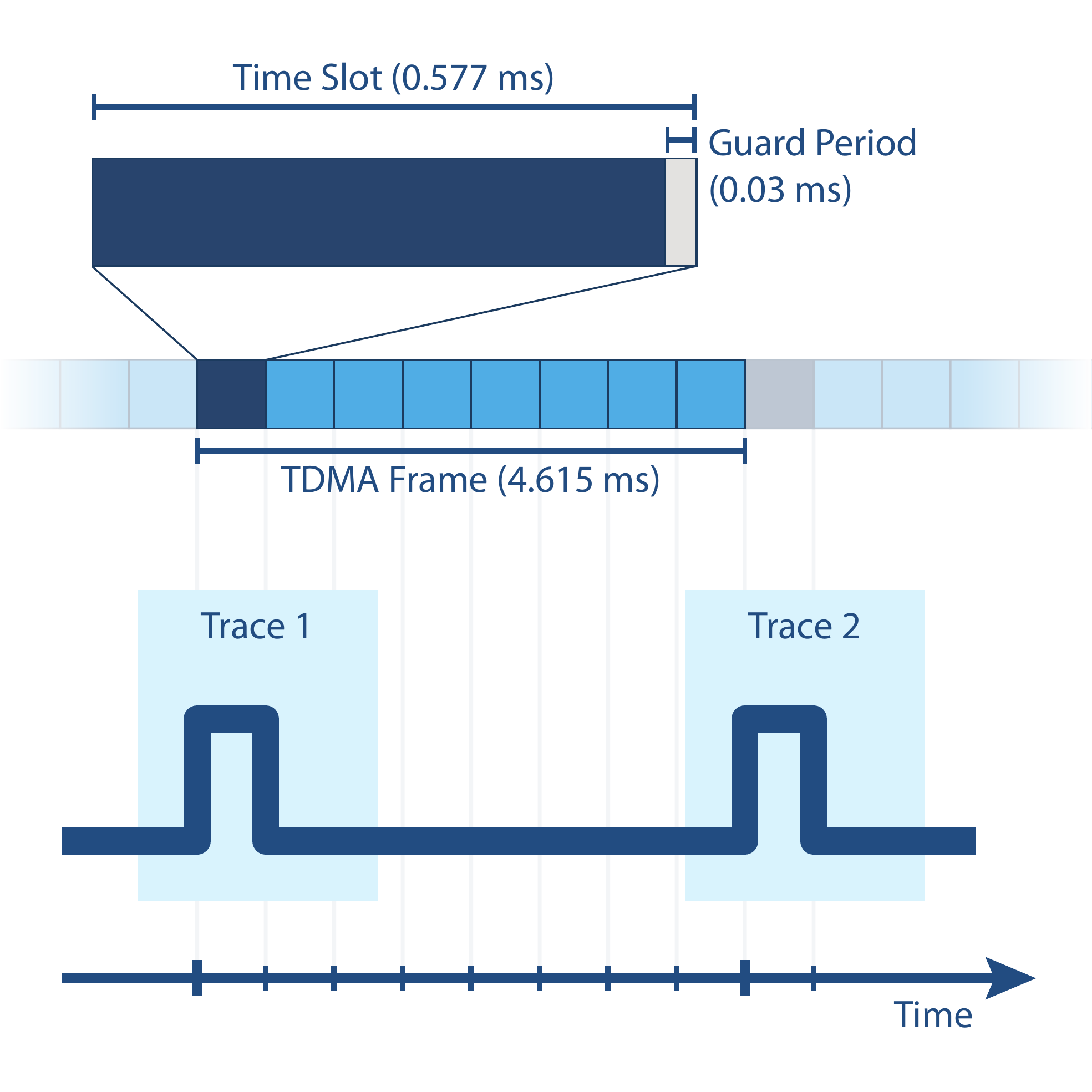}
    \caption{\textit{Left:} Illustration of a pileup-on-tail outside the time window (\pot) event. The pulse shape of Trace 2 is distorted by the tail of Trace 1. Two traces are saved with a timestamp difference \dtch. \textit{Right:} Schematic of the time trace of a GSM signal triggered in one detector channel. The time structure of the corresponding GSM signal is indicated above}
    \label{fig:pulseschemes}
 \end{figure*}
 
The majority of triggered traces are \ho events with an energy-independent pulse shape as shown in \cref{fig:pulse} (right) and \cref{fig:pulseshapes}a, and a statistically distributed \dtch depending on the activity of the particular detector channel. In addition, there are traces from various sources which, if not recognised and eliminated, can distort the spectrum. They can be divided into pileup originating from \ho, and spurious signals from external sources.
\subsection[Holmium-163 pileup]{\ho pileup}
The pulse shape of a \ho event can be distorted if a second event in the same detector occurs within a relatively short time interval \dtch. For a time difference larger than the time window $\dtch > \SI{2}{\milli\second}$, individual traces are triggered and saved for the two pulses as illustrated in \cref{fig:pulseschemes} (left). The pulse shape of the second trace is distorted by the tail of the previous pulse, as shown in \cref{fig:pulseshapes}b. For \dtch smaller than a given value, which depends on the time profile of the signals, this distortion can result in an incorrect reconstruction of the amplitude. Events of this kind are referred to as ``\textbf{p}ileup-on-tail \textbf{o}utside the \textbf{t}ime window'' (\pot). The distortions can be omitted by selecting traces with sufficiently high values of \dtch, as discussed in \cref{sec:holdoff}.

If $\dtch < \SI{2}{\milli\second}$, only one trace is saved with a trigger time corresponding to the occurrence of the first event. For very short time differences $\dtch \sim \risetime$, the pileup of two events with energies $E_1, E_2$ cannot be distinguished from the trace of one event with an energy of $E \simeq E_1+E_2$. This unresolved pileup is examined in more detail in \cref{sec:energy_dep} and can be taken into account statistically.

Events with $\dtch < \SI{1.57}{\milli\second}$ but $\dtch \gtrsim \risetime$ are referred to as ``\textbf{p}ileup-on-tail with both signals \textbf{i}nside the \textbf{t}ime window'' (\pit). For these events, the tail of the first pulse deviates strongly from the regular pulse shape (See \cref{fig:pulseshapes}c). A template fit in which a reference pulse is scaled to the trace would provide a false amplitude. Identifying and discarding such events is possible by means of a larger \chitworeduced value of the fit, which is described in \cref{sec:psa}.

\subsection{External Spurious Signals}
    \paragraph{Particle Background:}
        Natural radioactivity and cosmic muons can produce events in the energy range of the \ho spectrum. Not all these events can be distinguished from \ho events by means of their pulse shape (see \cref{fig:pulseshapes}d). Background suppression measures and a background model for the ECHo-1k setup are therefore of major importance to reliably analyse the endpoint region of the \ho spectrum \cite{Goe21}. Coincident signals could arise from secondary particles generated by muons interacting in surrounding materials or from muons passing through a pixel and the substrate. Along this line, a search for coincident events among different detector channels allows to identify events part of these muon related events.
    
    \paragraph{Mobile Phone Signal:}
        We observed that mobile phone signals transmitted with the Global System for Mobile communication (GSM) \cite{GSM} can couple into our readout system and generate triggered traces (see \cref{fig:pulseshapes}e). The detailed underlying mechanism for this coupling is still under investigation. The time structure of a GSM signal is partitioned into time division multiple access (TDMA) frames with a duration of $\SI{120}{\milli\second}/26 \simeq \SI{4.615}{\milli\second}$. Each TDMA frame consists of eight equal time slots, each of which can contain a burst of data. Normally, a user is assigned to one of these time slots, which can cause repeating signals to be triggered with a period of $\sim$\,\SI{4.615}{\milli\second}, as illustrated in \cref{fig:pulseschemes} (right). In consideration of the respective guard periods one expects a burst duration of \SI{0.5465}{\milli\second} for a normal burst (i.e. digitised voice data) and \SI{0.3210}{\milli\second} for an access burst (i.e. communication to the base station).
        
    \paragraph{Triggered Noise:}
        In addition to the sources mentioned above, miscellaneous temporary electromagnetic spurious signals can couple into the readout chain and create a false triggered signal. One example for this are small fluctuations in the power grid. Typically, those signals are characterised by a quickly repeating time signature and an anomalous shape of the trace (see \cref{fig:pulseshapes}f).

\section{Data Reduction}
Various methods have been studied to filter spurious events in microcalorimeters. The main objectives are mitigating nonlinearity of the detector output, reconstructing single events from pile-up events, lowering the threshold for unresolved pile-up, and detecting outliers. Most approaches are based on either (modified) optimal filtering techniques \cite{Shank2014,Wulf2016} or on principal component analysis \cite{Busch2015,Alpert2015,Fowler2015,Fowler2019,Borghesi2021}. While these methods show promising results, in this study we focus on arithmetically simple approaches as we aim for a fast \textit{online} data reduction. For this, the developed algorithms have been tested \textit{offline} with an available dataset first, and will be implemented into the data readout scheme in the future.

\begin{figure*}[!t]
    \centering
    \includegraphics[width=\textwidth]{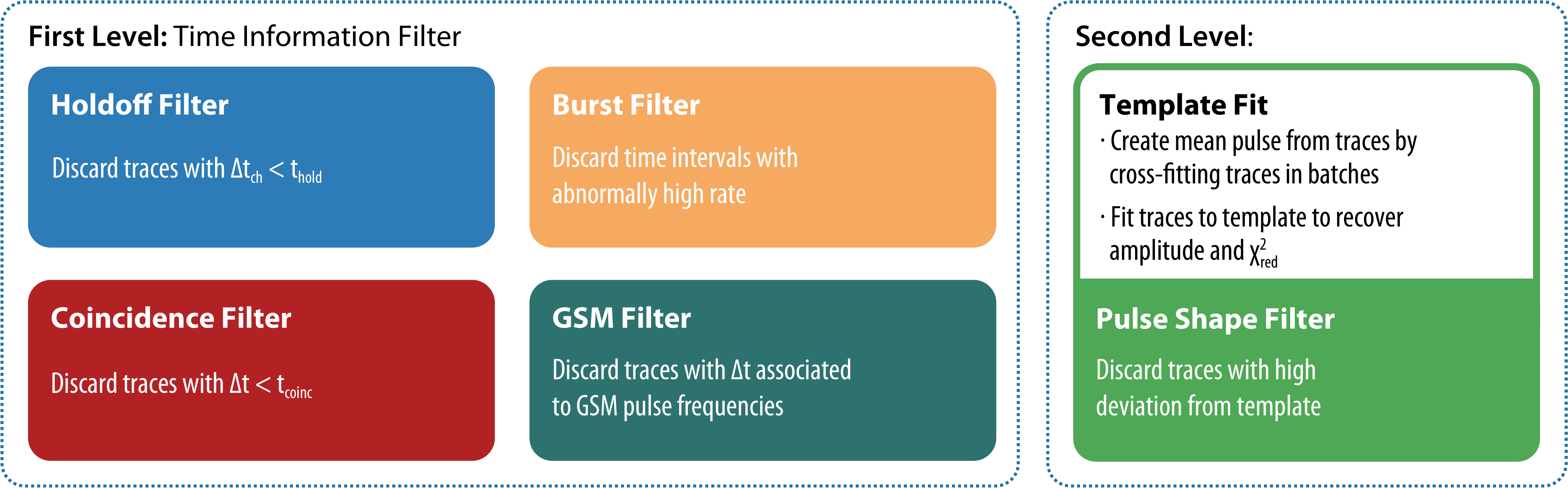}
    \caption{Schematic of the two-level offline data reduction algorithm described in the text}
    \label{fig:dataReduction}
\end{figure*}

The presented offline data reduction algorithm comprises two levels as illustrated in \cref{fig:dataReduction}. On the first level, only the trigger time information of the acquired raw traces is used to discard \pot events and external spurious signals in an energy-independent way. On the second level, the remaining data filtered by the first level filter are further analysed based on their time profile to discard remaining spurious signals. A template pulse is automatically generated by averaging \ho traces of the MI-line. Traces that deviate strongly from the template are then discarded.

\subsection{Time Trigger Information Analysis}
    The \textit{\tifilter} is defined as the logical \verb|AND| of four independent subfilters, each applied to the raw data. Thus, if a trace is discarded by at least one of the four subfilters, it is discarded by the \tifilter. In the following, the aim and implementation of each of the subfilters is described. The holdoff and burst subfilters are performed channel by channel, with \dtch being analysed. The coincidence and GSM subfilters are done globally, analysing \dt.
    
    \subsubsection{Holdoff Subfilter}
        \label{sec:holdoff}
        The aim of the holdoff subfilter is to discard \pot events. For this, traces are removed that fulfil
        \begin{align}
            \dtch < \tholdoff .
        \end{align}
        The holdoff time \tholdoff is fixed based on the time profile of a typical \ho signal such that the distortion of a pulse with $\dtch = \tholdoff$ is sufficiently small to ensure correctly reconstructed amplitudes at all energies. 
        The holdoff time is determined in dedicated characterisation measurements prior to the actual experiment run. For this purpose, traces are acquired over a large time window up to the point where the temperature pulse recovers its initial voltage value. This is done for both AC and DC coupled signals.
        
        This subfilter only removes the trace on the tail, i.e. the pulse occurring at a time interval $\dtch<\tholdoff$ after a previously triggered pulse in the same detector channel. For \echo run 24 with AC-coupled signals, a value of \tholdoff = \SI{15}{\milli\second} was determined.
    
    \subsubsection{Burst Subfilter}
        In order to discard any traces from quickly repeating triggered noise, the burst subfilter identifies time intervals with an abnormally high trigger rate. The subfilter is applied channel by channel, since noise usually does not couple identically in all detector channels.
        
        The timestamps of traces of each detector channel are binned with a bin width \binwidth. The expected number of events from \ho decay per bin is then given by
        \begin{align}
            \nexcpect = \activitych \binwidth,
            \label{eq:nexpect}
        \end{align}
        where \activitych is the \ho activity in the corresponding detector channel known from detector characterisation. A bin that contains quickly repeating triggered noise will exhibit a number of counts that strongly exceeds the expected value \nexcpect of the otherwise dominant \ho events. The degree of deviation from the expected value can be expressed in terms of the statistical uncertainty of \nexcpect, which for a Poissonian distribution is given by its standard deviation $\sigma = \sqrt{\nexcpect}$. The traces within a bin are discarded if the number of counts exceeds $\nexcpect + 4\sigma$. If a bin fulfils this criterion, it is referred to as a seed bin. For the two neighbouring bins of a seed bin, the threshold for the bins to be discarded is lowered to $\nexcpect + 2\sigma$. This ensures that no fragments of a burst are missed due to binning.
        
        Two complementing burst subfilters are implemented, one optimised for faster bursts and the other for slower bursts. The difference lies in the way the bin width is defined. For fast bursts it is chosen such that \nexcpect $=1$, i.e.
        \begin{align}
            \binwidth = \activitych^{-1}.
        \end{align}
        This corresponds to the shortest bin width that can reasonably be defined.
        
        In order to be sensitive to noise triggered with a frequency down to \fnoise, the bin width of the second burst subfilter is defined in a way that $\fnoise  \binwidth = 4\sigma$. With the definition of $\sigma$ and \cref{eq:nexpect}, this condition is fulfilled for
        
        \begin{align}
            \binwidth = 16 \frac{\activitych}{\fnoise^2}.
        \end{align}
        The burst subfilter is then defined as the logical \verb|AND| of the decisions made with both methods. Thus, if a trace is discarded by at least one of the two methods, it is discarded by the burst subfilter.
    \begin{figure*}[!ht]
        \centering
        \includegraphics[height=0.57\textwidth]{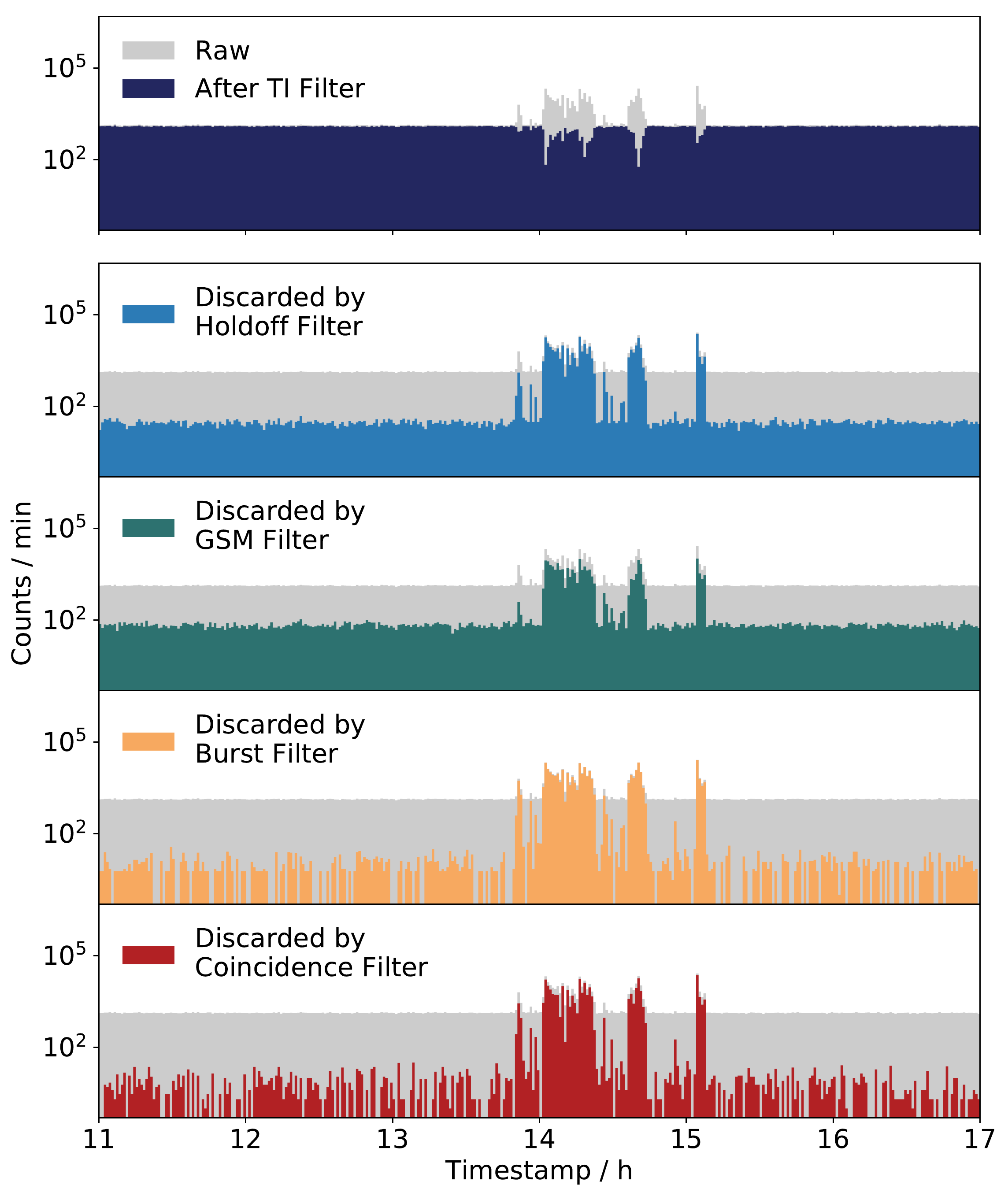}
        \includegraphics[height=0.57\textwidth]{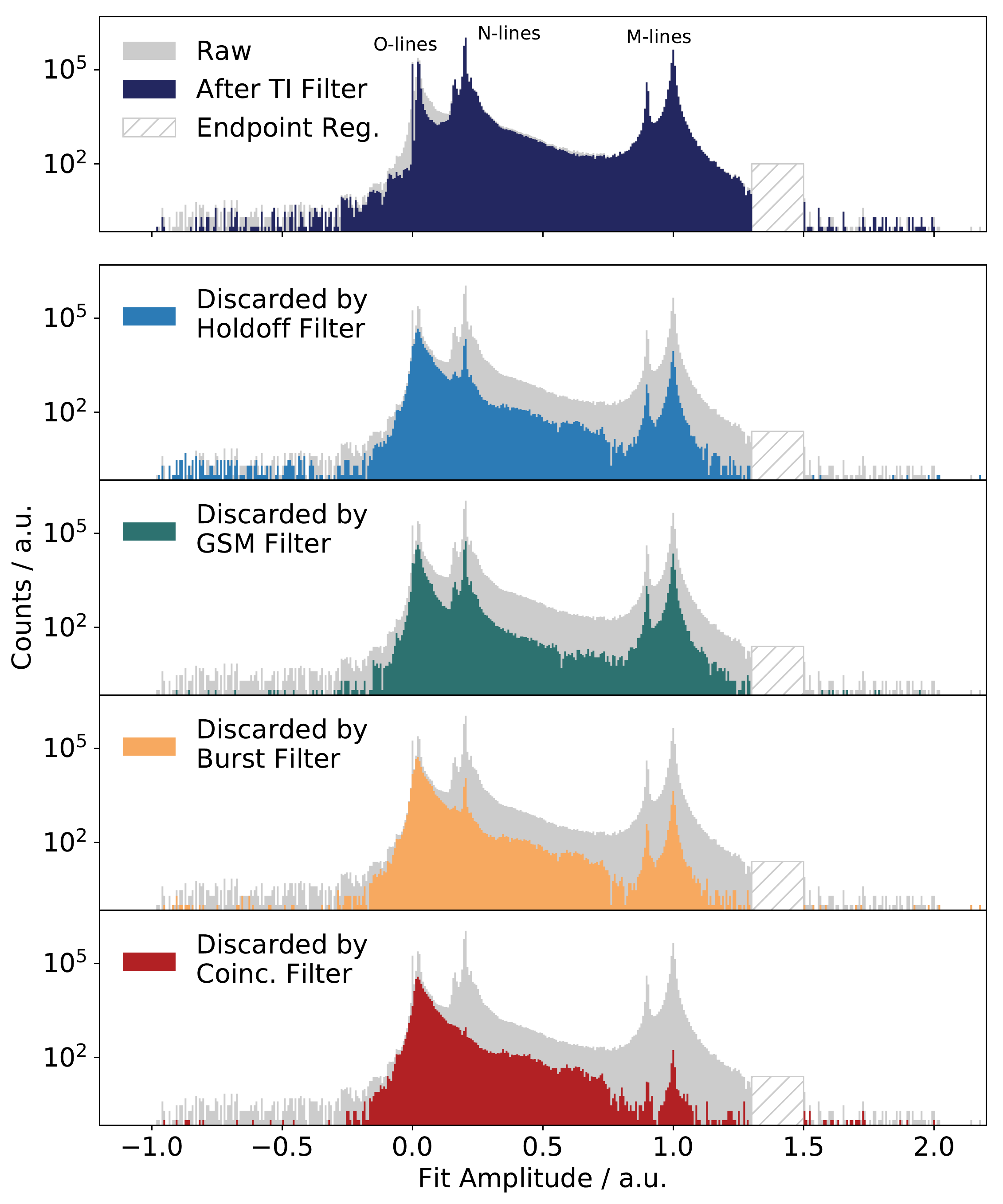}
        \caption{Histograms to illustrate the influence of the \tifilter, broken down by the four independent subfilters. \textit{Left:} Histograms of the number of acquired traces per minute. From a two-day dataset of ECHo run 24, a six hour sample containing high levels of noise between the timestamps \SI{13.5}{h} and \SI{15.5}{h} is shown. \textit{Right:} Histograms of the fit amplitude. In all ten panels, the raw histogram prior to any filters is plotted in grey. In the two top panels, the histograms after application of the \tifilter (TI filter) can be seen. The panels below present the histograms of traces discarded by the individual subfilters}
        \label{fig:TIFilter_summary}
    \end{figure*}     
    
    \subsubsection{Coincidence Subfilter}
        For an activity of \SI{1}{\becquerel} per detector channel, which is typical for ECHo-1k, coincidence among different detector channels on a microsecond timescale due to \ho has a low probability. Muon-induced events or certain electromagnetic signals in turn often cause triggered events in multiple detector channels at the same time. Thus, discarding coincident events altogether is an efficient way to reduce spurious signals. Traces that fulfil 
        \begin{align}
            \dt < \tcoincidence
        \end{align}
        as well as the corresponding previous traces are considered coincident.
        
        The coincidence time \tcoincidence can be defined by the time response of the signal. In the discussed datasets, the time response is governed by the gain bandwidth product (GBP) of the amplification circuit. One usually obtains an effective time resolution \risetime of a few hundred nanoseconds. For muon related events, $\dt < \risetime$ holds. However, for electromagnetic signals that couple into the readout scheme of multiple detector channels, time differences \dt up to a few microseconds have been observed. Therefore, a conservative coincidence time of $\tcoincidence = \SI{8}{\micro\second}$ is used for \echo run 24.

    \subsubsection{GSM Subfilter}
        This subfilter is implemented to specifically reject triggered GSM phone signals. For this, characteristic \dt values associated with GSM signals are defined. Besides integer multiples of the duration of a TDMA frame, this includes the burst duration of a normal burst and an access burst. The burst duration can appear in the data stream when the rising and falling edges of a burst are triggered in different detector channels. Traces with a relative \dt within a \SI{\pm20}{\micro\second} interval around one of these characteristic \dt values are discarded.
        
        In principle, there is an infinite number of characteristic time differences \dt that can be associated when considering all integer multiples. In practice however, a maximum value of \dt is defined according to the total activity of the chip such that the probability that two triggered GSM signals separated by \dt are not interrupted by a \ho signal is $10\;\%$.

    \subsubsection{Application of the Time Information Filter}

        The first level filter is applied to a dataset acquired with 34 implanted pixels of one ECHo-1k chip during two days of run 24  with a total activity of $\sim$\,\SI{25}{\becquerel}. In the presented data reduction routine, the template fit as described in \cref{sec:templatefit} is only performed for data that passes the \tifilter. For this two-day dataset however, amplitudes are obtained for all data in order to illustrate the working principle of the \tifilter as well as to assess its efficiency (\cref{sec:TIFilter_Assessment}).

        In \cref{fig:TIFilter_summary}, the number of acquired and discarded traces per minute (left) and the fit amplitudes of acquired and discarded traces (right) are shown for 18 detector channels of the two-day dataset acquired with an ECHo-1k chip. For both plots, the top panel shows the corresponding histograms after applying a \tifilter consisting of all four subfilters. The lower panels show the histograms of discarded traces broken down by by the individual subfilters. In all ten panels, the histogram of acquired raw traces is shown in grey for comparison. For most of the acquisition time, the number of events acquired per minute is constant, as mostly \ho events are triggered. Between the timestamps of \SI{13.5}{h} and \SI{15.5}{h}, the number of counts increases by up to an order of magnitude. After applying the \tifilter, the number of counts per minute within this time interval drops below the average undisturbed value, while the undisturbed region is barely affected. During the period of a high count rate, the number of discarded traces increases strongly for all subfilters.
        
        In the histograms of discarded fit amplitudes shown in \cref{fig:TIFilter_summary} (right, lower panels), one can see the reconstructed amplitudes of discarded background traces as well as a component of falsely discarded \ho traces. The spectrum of traces discarded by the coincidence subfilter, which can be seen in the bottom panel, is a background spectrum with only few discarded \ho traces. It is characterised by a strong increase of counts towards low fit amplitudes, particularly below the NI-resonance. It is important to note that the fit amplitudes of background signals cannot necessarily be translated to an energy scale as is the case with \ho signals.
        \begin{figure}[!t]
             \centering
             \includegraphics[height=0.373\textwidth]{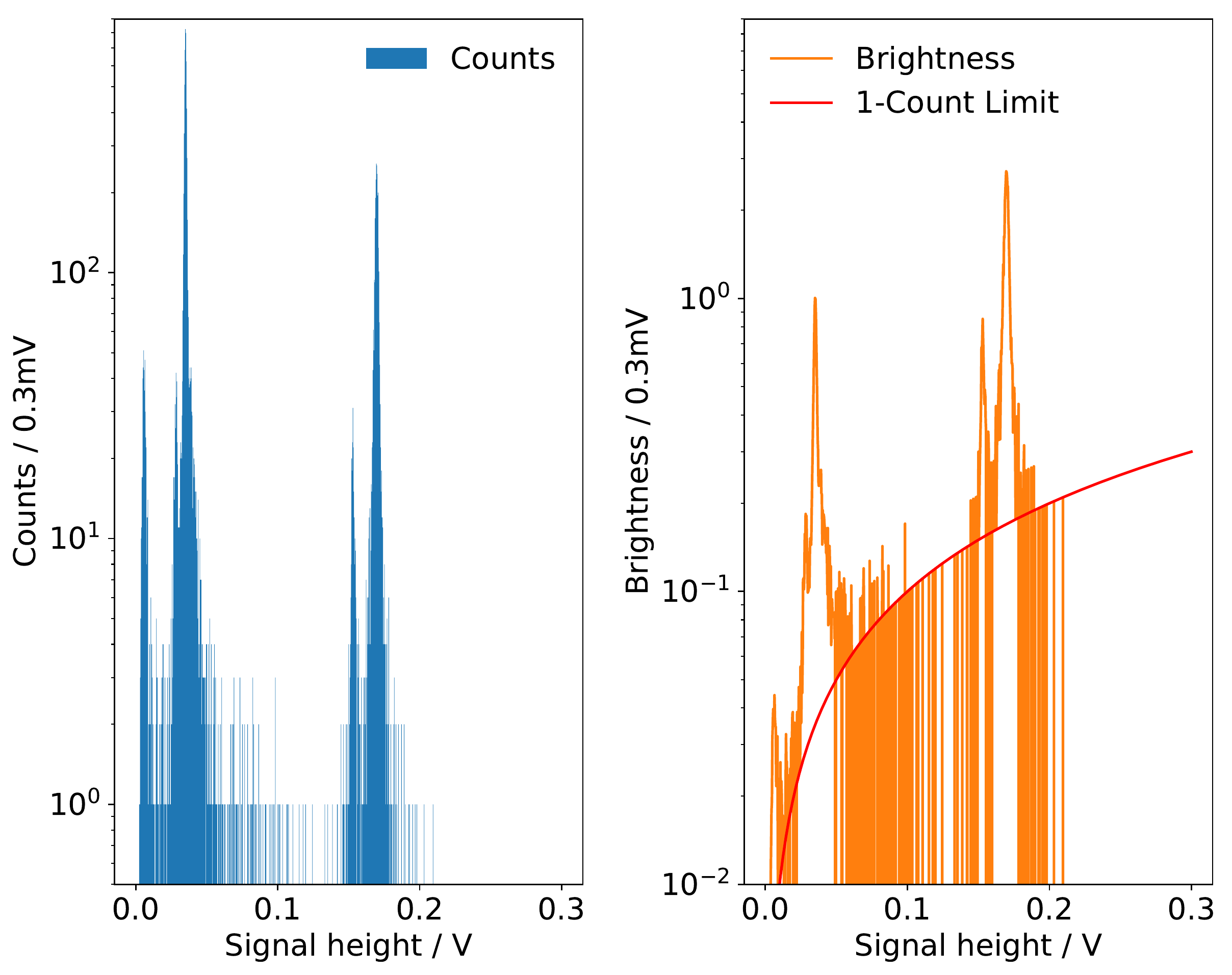}
             \caption{Representation of the signal heights of the first 10000 raw traces of a single pixel. \textit{Left:} Histogram of the signal heights \textit{Right:} Brightness of the traces calculated via \cref{eq:brightness}. One can clearly see that the MI-line is the brightest. This holds for much higher signal heights as well, as highlighted by the 1-count limit in red. The number of bins (1000) is chosen simply for optical clarity and only slightly influences the position of the maximum}
             \label{fig:Brightness}
        \end{figure}
    \subsection{Pulse Shape Analysis}
        \label{sec:psa}
        The aim of the second level of data reduction is to recognise and eliminate \pit as well as time-uncorrelated noise traces. For this, a mean trace (template) is generated for each pixel. All traces that have passed the first filter will then undergo a template fit with the obtained template. The goodness of fit parameter \chitworeduced is calculated, which provides a measure for how well each trace can be scaled to the template. This is used to define the second level filter.

        \begin{figure}[!t]
            \centering
            \includegraphics[height=0.379\textwidth]{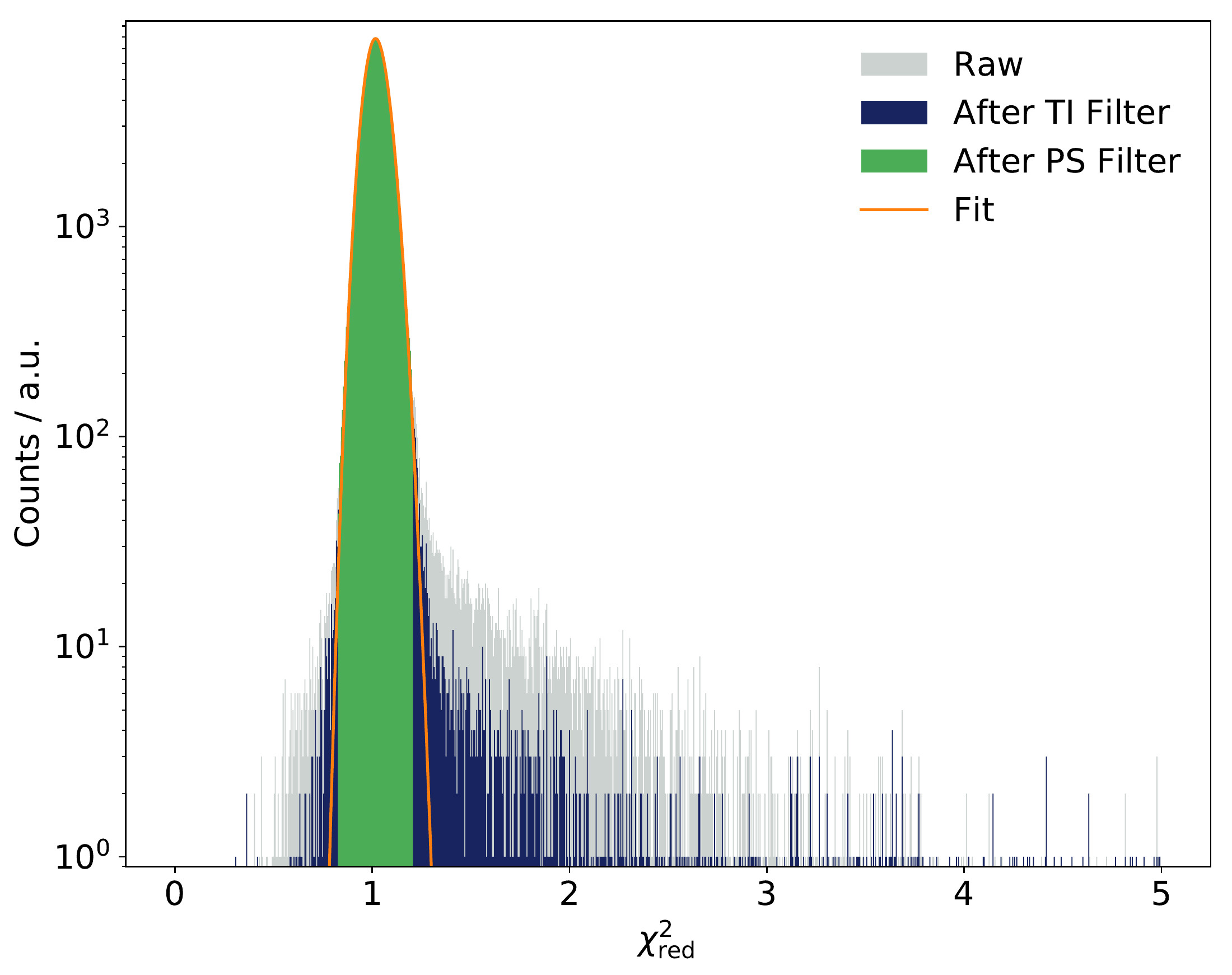}
            \caption{Histogram of \chitworeduced of a single detector channel. The plotted range contains $\simeq \SI{99.7}{\%}$ of all raw events. The filtered region after the \psfilter still encompasses $\simeq \SI{97.6}{\%}$ of all raw events. Also depicted is the skewed Gaussian fit from which the \psfilter is calculated}
            \label{fig:Chi2Distribution}
        \end{figure}
    
    \subsubsection{Automated Template Generation}
        \label{sec:template}
        
        It is apparent from \cref{fig:pulse} (right) that the shapes of the traces from a single energy deposition in the detector are energy independent. Therefore, it is possible to build a discrimination scheme based on the deviation of traces from the general shape called the \textit{template}. In order to process the vast amount of data acquired for the ECHo-1k experiment, an automated process for generating templates has been developed. To ensure a high reliability of the final pulse shape, the template is generated by averaging a large number of individual traces belonging to the MI-line of the \ho spectrum. In fact, MI-events already have a very good \textit{figure-of-merit} \figureofmerit at the level of single events, defined as:
        \begin{equation}
            \figureofmerit = \frac{\signalheight}{\pretriggernoise},
            \label{eq:figureofmerit}
        \end{equation}
        where \signalheight and \pretriggernoise are the height of the template signal (the average of the 10 samples after the maximum) and the standard deviation of the noise of the pre-trigger samples (\textit{pre-trigger noise}), respectively. A high \figureofmerit corresponds to a signal shape relatively undisturbed by noise. As an example, compare a pulse from the MI-line ($\figureofmerit \sim 100 - 400$) to a pulse from the OI-line ($\figureofmerit \sim 10 - 40$), as shown in \cref{fig:pulse}.\\

        To reach higher \figureofmerit values, multiple traces have to be averaged to form the template. By averaging $N$ traces, the pre-trigger noise of the resulting template is reduced by a factor of $\counts^{-1/2}$. Ideally, the traces should already have a high signal height. In order to maximise the \figureofmerit of the template, the best approach is to select a region of the available spectrum with a high fraction of \ho signals compared to traces from various other sources as detailed in \cref{sec:signalFamilies}. This is the case for energies close to the main resonances of the \ho spectrum. Also, selecting only traces from a single resonance as opposed to multiple ones ensures that \figureofmerit increases reliably.
        
        The MI-resonance was found to be best suited for template generation based on the following approach: first, a histogram of the signal heights in \si{\volt} as provided by the acquisition software of the first few traces (${\simeq 10000}$) is generated, as shown in \cref{fig:Brightness} (left). Then, the \textit{brightness} \brightness is calculated per bin, defined by
        \begin{equation}
            \label{eq:brightness}
            \brightness = \signalheight \sqrt{\intensity},
        \end{equation}
        where \intensity is the number of counts in each bin. The result can be seen in \cref{fig:Brightness} (right). Summing all traces of the region with the maximum \brightness will result in an average trace with the maximum \figureofmerit. For the \ho spectrum, the brightest line is the MI-line. Hence, as stated above, it is the ideal candidate for template generation.
        
        The signal height of the MI-line is located via a peak detection algorithm based on a continuous wavelet transform as implemented in the Python package SciPy \cite{SciPy01}, performed on \brightness.

        The last step is to iteratively read in traces with a signal height that is within \SI{1}{\%} tolerance of the mode of the MI-line in small batches of 200. One can then filter those traces with \pit and other defects by calculating their pairwise quadratic differences in a vectorised manner, discarding those that deviate from the median quadratic difference by a factor $\geq 2$. The remaining traces can be averaged until an initially defined \figureofmerit for the template is reached. If the data is exhausted before reaching the intended \figureofmerit, this particular dataset is discarded from further analysis.
    
    \subsubsection{Template Fit Method}
        \label{sec:templatefit}
    
        Once the average MI-signal is generated, a template fit for all the traces which survived the first level filter is performed. The measure used to determine how well the shape of the traces agrees with the template is the reduced chi-square, defined as:
        \begin{equation}
            \chitworeduced = \frac{1}{f}\sum_{i=1}^f \left( s_i - \amplitude \theta_i - \offset \right)^2,
        \end{equation}
        where \amplitude and \offset describe the amplitude and offset of the trace $\vb{s}$ respectively, $\boldsymbol{\theta}$ is the template, and the sum runs over all $f$ elements of $\vb{s}$ and $\boldsymbol{\theta}$. \chitworeduced is then minimised with respect to \amplitude and \offset. The normalisation by $1/f$ (roughly the degrees of freedom of the fit) causes ${\chitworeduced \simeq 1}$ for a \ho signal and facilitates further evaluation and analysis.
    
        After all traces have been fit, a histogram of all \chitworeduced is generated as shown in \cref{fig:Chi2Distribution}, where in a typical measurement, $\simeq \SI{99.5}{\%}$ of all traces are inside the region of ${0 \leq \chitworeduced \leq 5}$. A skewed Gaussian distribution is fit to the histogram and the \textit{\psfilter} is defined such that all traces which lie outside the $\SI{99.73}{\%}$-region\footnote{This would be the $3\sigma$-region of a non-skewed Gaussian distribution.} of the skewed Gaussian are discarded.

\section{Assessment of the Data Reduction Algorithm}
In the following, the performance of the filters defined above is evaluated for a subset of ECHo-1k data. For the first level filter, the efficiencies to retain signal and reject background are estimated based on the two-day dataset acquired with an ECHo-1k chip with a total activity of $\sim \SI{25}{\becquerel}$ as well as using simulated \dt values. Hereinafter, the energy dependence of the second level filter is analysed based on simulated \ho traces.

\subsection{Assessment of the Time Information Analysis}
    \label{sec:TIFilter_Assessment}
    \subsubsection{\ho Selection Efficiency}
    The fraction of \ho events that are unaffected by the \tifilter, which we will simply call signal efficiency, is estimated for each subfilter individually. For the holdoff subfilter, by definition, the signal efficiency is $\epsilon_\mathrm{hold}^\mathrm{sig}=\SI{100}{\%}$. Even though not all discarded traces would cause a falsely reconstructed amplitude, it is crucial for them to be rejected in order to obtain an energy-independent filter. Hence, all traces discarded by this subfilter are considered a source of background.
    
    For the GSM subfilter as well as for the coincidence subfilter, \ho traces are randomly discarded if their time difference \dt lies in a region that is associated with mobile phone signal or coincident events respectively. The fraction of \ho events occurring within the time intervals related to GSM signals and the fraction of random coincidence of \ho events are obtained by applying the subfilters on simulated data with values of \dt distributed according to a total activity of $A = \SI{25}{\becquerel}$. One finds that the fraction of \ho events removed by applying the GSM subfilter is \SI{5}{\%} while in case of the coincidence subfilter the fraction of discarded \ho traces amounts to \SI{0.04}{\%}. This corresponds to signal efficiencies of $\epsilon_\mathrm{GSM}^\mathrm{sig} = \SI{95}{\%}$ and $\epsilon_\mathrm{coinc}^\mathrm{sig} = \SI{99.96}{\%}$ respectively. As expected, it can be seen in \cref{fig:TIFilter_summary} (right) that for the two-day data set, the number of \ho traces discarded by the GSM subﬁlter exceeds the number of \ho traces discarded by the coincidence subﬁlter by more than two orders of magnitude.
    
    For the burst subfilter, the effective off-time caused by the rejection of time intervals with abnormally high rates is determined for each detector channel. The number of falsely discarded \ho traces can be estimated from the product of activity and effective off-time for each detector channel. For the two-day dataset, the fraction of effective off-time to acquisition time ranges from $0.1\;\%$ for low-noise detector channels to $1.5\;\%$ for detector channels with strong coupling of mobile phone signals. The total fraction of \ho traces discarded by the burst subfilter is \SI{0.7}{\%} and thus $\epsilon_\mathrm{burst}^\mathrm{sig} = \SI{99.3}{\%}$.
    
    \subsubsection{Background Rejection Efficiency}
    The efficiency to reject signals which could contribute as background to the \ho spectrum is defined individually for the signal families specified in \cref{sec:signalFamilies}. To reject \pot events, the holdoff time is conservatively chosen such that the broadening of the spectral shape due to falsely reconstructed amplitude is negligible. Thus, a background rejection efficiency for \pot of $\epsilon_\mathrm{\pot}^\mathrm{bkg}=\SI{100}{\%}$ can be assumed.
    
     \begin{figure}[!t]
         \centering
         \includegraphics[width=\linewidth]{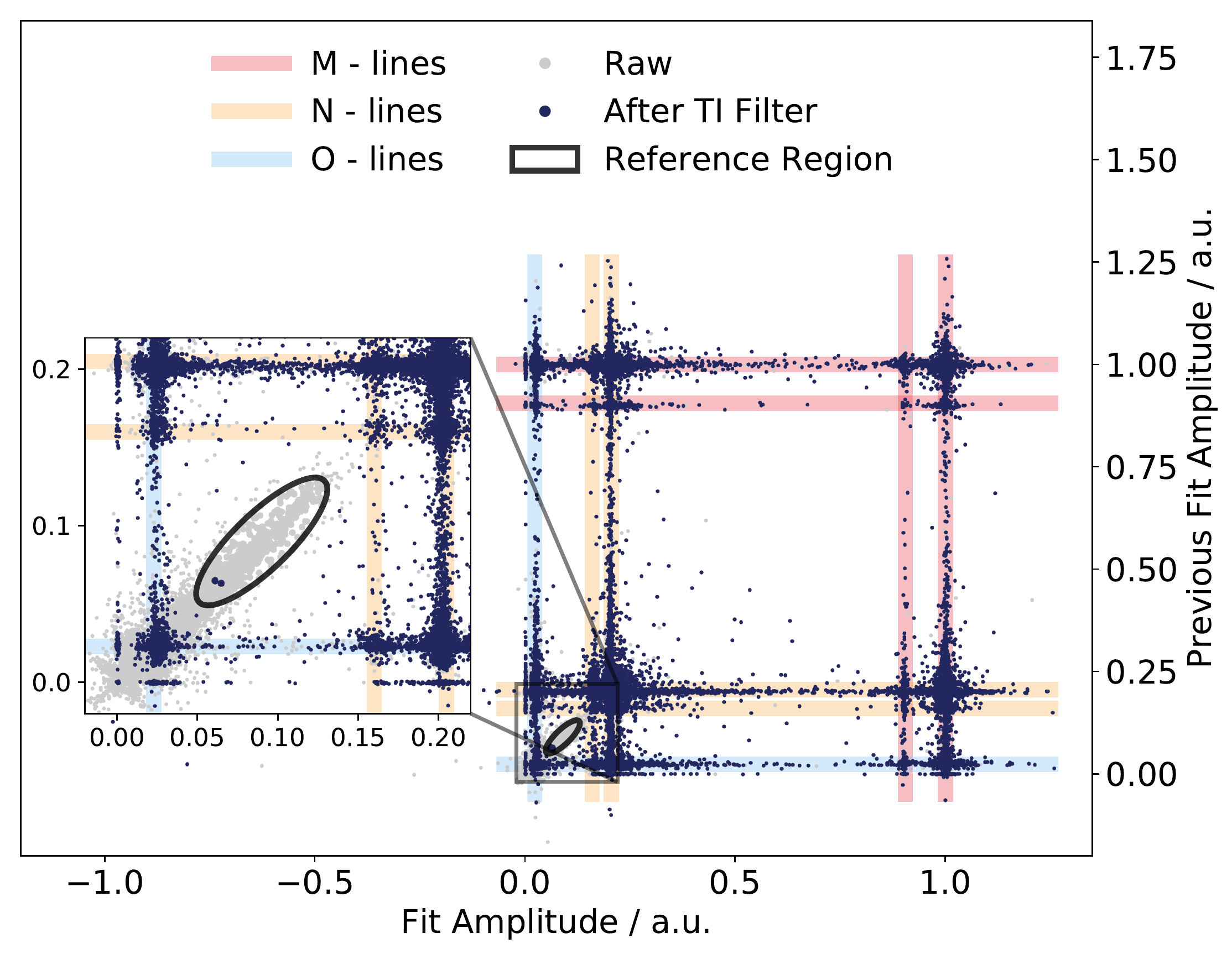}
         \caption{Fit amplitude of a trace versus fit amplitude of the preceding trace in the same detector channel before and after a \tifilter is applied. Data points from one detector channel of the two-day dataset are shown. \ho events are mostly located on lines parallel to the main axes corresponding to the individual resonances of the \ho spectrum. High densities appear at the intersections of two such lines. The inset shows a magnification of the region between the N-resonances and O-resonances together with a structure of correlated amplitudes originating from mobile phone signals. The ellipse-shaped reference region is used to estimate the background rejection efficiency for mobile phone signals. Data points within the ellipse are magnified for better readability. One can see that nearly all of the signals within the ellipse are discarded by the \tifilter while most of the \ho signals are unaffected}
         \label{fig:Correlation_Space}
     \end{figure}
    
    To estimate the background rejection efficiency for mobile phone signals, we define a reference region in a parameter space with particularly high background-to-signal-ratio. The background rejection efficiency can then be estimated based on the fraction of events within this region that are discarded. A good way to separate GSM events from \ho events is to plot the amplitude of the acquired traces against the amplitude of the preceding trace for all acquired traces in one detector channel. As indicated in \cref{fig:Correlation_Space}, \ho events are mainly distributed in regions parallel to the main axes. Triggered noise was found to accumulate along the diagonal through the origin with unit slope. The correlation of fit amplitudes of subsequent mobile phone signals is not surprising, since triggered mobile phone signals are typically characterised repeating signals with the same shape. The reconstructed amplitudes of mobile phone signals are well below the M-resonances and in the continuum region between two resonances, the \ho event rate is suppressed by several orders of magnitude compared to the region around the resonances. Thus, an ellipse-shaped reference region can be defined, centred between the NII- and OI-resonances, as drawn in \cref{fig:Correlation_Space}, which has a particularly high background-to-signal-ratio. In this way, traces contained within the ellipse can be considered a pure sample of background signals. For the two-day dataset, only 77 out of 44602 events within the ellipse  are not discarded by a \tifilter consisting of all four subfilters. The efficiency is estimated by the fraction of events within the ellipse that are discarded by the \tifilter $\epsilon_\mathrm{phone}^\mathrm{bkg}=\SI{99.826\pm0.020}{\%}$. The error is obtained from the variance of a Binomial distribution\footnote{This naive approach does not hold for efficiencies close to 0 or 1. However, using the Bayesian approach discussed in \cite{Casadei2012} only weakly affects the result.}.
    
    As for particle background, only signals initiated by atmospheric muons can be tackled with the \tifilter. Such traces can be discarded if a coincident signal in multiple pixels is produced. The efficiency of rejecting muon-induced events by means of the coincidence subfilter can be estimated from an acquisition with an active muon veto installed around the dilution refrigerator. The background rejection efficiency for muon-induced events is the ratio of pixel-pixel-veto coincidences and pixel-veto coincidences, i.e. the fraction of muon-induced signals that produce a signal in at least two pixels. In \cite{Goe21}, a measurement with muon veto was described for 64 pixel-days. A total of \SI{242\pm 20}{} pixel-veto coincidences and \SI{194\pm 12}{} pixel-pixel-veto coincidences were measured. Thus, one can derive $\epsilon_\mathrm{muon}^\mathrm{bkg}=\SI{80\pm8}{\%}$.
    
    \subsubsection{Necessity of the Individual Subfilters}
    \pot events are discarded efficiently by the holdoff subfilter and muon related background is tackled by the coincidence subfilter. Since these signal families are each addressed by only one subfilter, the use of both the holdoff subfilter and the coincidence subfilter is essential. This also implies that the combination of different subfilters does not improve the respective rejection efficiencies for traces from these signal families.
    
    To evaluate the benefit of the additional use of a burst subfilter and a GSM subfilter to specifically reject mobile phone induced noise, the influence of these subfilters on the background rejection efficiency $\epsilon_\mathrm{phone}^\mathrm{bkg}$ is investigated. If instead of all four subfilters only the holdoff subfilter and the coincidence subfilter are applied, 226 instead of 77 out of a total of 44602 mobile phone signals within the ellipse are not discarded. If in addition to these two subfilters the burst subfilter is applied, 81 traces remain undiscarded. In the case of using the GSM subfilter in addition to the holdoff subfilter and the coincidence subfilter, 103 traces in the ellipse are not discarded.
    
    Even though the improvement due to the additional subfilters seems to be minor for this dataset, the burst subfilter in particular should always be applied. The signal efficiency for this subfilter is already high and increases even further the lower the noise level of the acquisition. In addition, the burst subfilter is sensitive to abnormally high trigger rates that only occur in one detector channel, and even to rather low noise frequencies that cannot be resolved by any of the other subfilters. Applying a \tifilter without the GSM subfilter, the background rejection efficiency $\epsilon_\mathrm{phone}^\mathrm{bkg}$ is still above \SI{99.8}{\%} and the signal efficiency is dominated by $\epsilon_\mathrm{burst}^\mathrm{sig} = \SI{99.3}{\%}$.
    
    With a signal efficiency of only \SI{95}{\%}, the GSM subfilter is an expensive filter in terms of discarding good data. Furthermore, the reconstructed energy of GSM signals is well below \qec and thus won’t affect the spectral shape close to the endpoint. In the two-day dataset of ECHo-1k, the additional application of this particular subfilter shows no advantage over the sole use of the other three subfilters. In future runs, the coupling of GSM signals will be reduced by improving screening of the read out components. For analyses of the low energy part of the spectrum however, where background levels increase, as well as for acquisitions with high levels of triggered mobile phone signal, this subfilter can be of relevance.

    \subsection{Assessment of the Pulse Shape Analysis}
    \begin{figure*}[!ht]
         \centering
         \includegraphics[width=\textwidth]{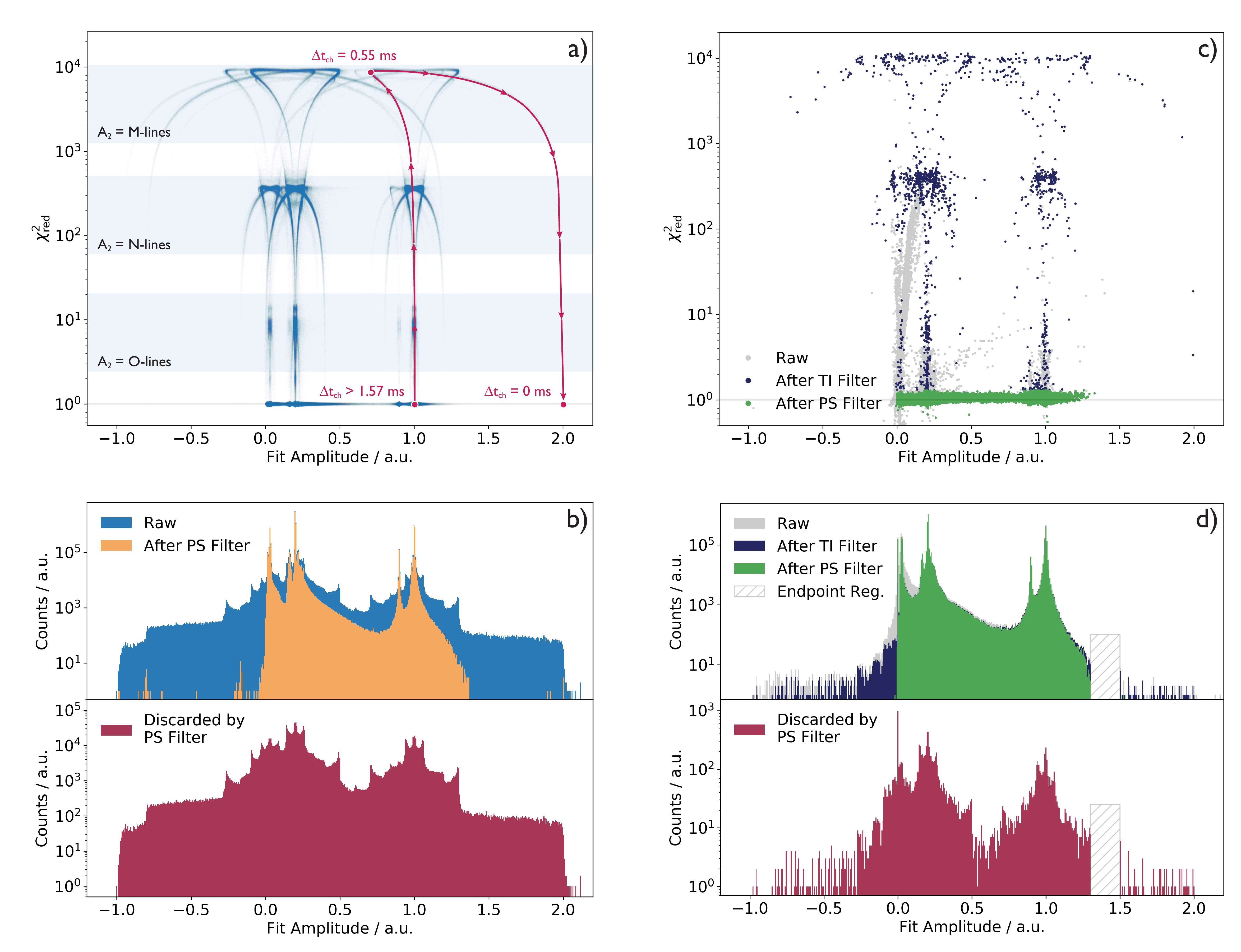}
         \caption{Simulated (a, b) and ECHo-1k data (c, d). 
         \textbf{a)}  Fit amplitude vs. \chitworeduced  scatter plot of simulated events. An exemplary path of decreasing \dtch for $A_1=A_2=1$ and $\Pi=+1$ is indicated in red. The location of the maximum of the arc-shaped structures is determined by the amplitude of the pulse on the tail $A_2$. 
         \textbf{b)} Histogram of the fit amplitude of simulated events before and after applying the \psfilter. The large fraction of pileup is caused by the truncated \dtch distribution used for the simulation. The structure in the histogram of traces discarded by the \psfilter is discussed in the text.
         \textbf{c)} Fit amplitude vs. \chitworeduced  scatter plot for one detector channel of the two-day ECHo-1k dataset. The arc-shaped structures become apparent after applying the \tifilter. 
         \textbf{d)} Histogram of the fit amplitude of the ECHo-1k dataset with 18 detector channels where the endpoint region is blinded. The slight asymmetry of spikes around the main \ho-lines is caused by detector channels with an asymmetric activity in the two pixels. The spike at fit amplitude 0 corresponds to triggered baselines. Note that the scale on the x-axis is the same for all plots
         }
         \label{fig:amplitude_chi2}
     \end{figure*}
    To assess the pulse shape analysis, the template fit described in \cref{sec:psa} is performed on a set of simulated \ho data. The aim is to find the sensitivity of identifying \pit events as a function of time difference and energies of two subsequent events. Traces of \pit as well as regular \ho events are simulated with amplitudes, timestamps and polarities randomly drawn from corresponding distributions.
    
    \subsubsection{Simulation of pileup-on-tail with both signals inside the time window}
    For the simulation of signals with \pit, $10^7$ events are generated, each with an amplitude of the triggered pulse $A_1$, an amplitude of the subsequent pulse $A_2$, a time difference to the subsequent pulse \dtch and a relative sign of the polarity to the subsequent pulse $\Pi$. The corresponding values are drawn randomly from the expected probability distributions of the parameters. For the two pulse amplitudes, this distribution is the theoretical \ho spectrum \cite{Bra20} normalised by its area. The values of \dtch are drawn from an exponential distribution $\propto \exp(-A \dtch)$ with activity $A=\SI{1}{\becquerel}$. Integer multiples of $\SI{128}{\nano\second}$ are allowed, which corresponds to the time difference between two sampled data points typically used for acquiring \echo data, as described in \cref{sec:Introduction}. For $\Pi$, a discrete distribution $P(\Pi=-1)=P(\Pi=+1)=0.5$ is employed. Normal distributed noise $\boldsymbol{\mathcal{N}}$ is generated with a constant \figureofmerit for a pulse height corresponding to the MI-line of the \ho spectrum. In the following, $\figureofmerit=300$ is used, which is a typical value for ECHo-1k data.
        
    A simulated pileup trace $\boldsymbol{PU}(A_1,A_2,\dtch, \Pi)$ is then generated according to
    \begin{align*}
        \boldsymbol{PU}(A_1,A_2,\dtch, \Pi) = A_1 \boldsymbol{\theta}(t_\mathrm{shift}=0) \\+ \Pi A_2 \boldsymbol{\theta}(t_\mathrm{shift}=\dtch) \\+ \figureofmerit^{-1}\boldsymbol{\mathcal{N}}(\mu=0,\,\sigma^{2}=1)
    \end{align*}
    where $\boldsymbol{\theta}(t_\mathrm{shift})$ is a template pulse (see \cref{sec:template}) shifted in time by $t_\mathrm{shift}$.
        
    For a time difference larger than $\dtch\simeq\SI{1.57}{\milli\second}$, the rising edge of the subsequent pulse lies outside the time window and thus the pulse shape of the initial event is not affected. For this reason, the distribution of \dtch is truncated at $\dtch=\SI{10}{\milli\second}$ for the simulation.
        
    In total, $10^7$ events are generated. The number of \pit events simulated with the truncated range of \dtch is equivalent to the number of \pit events from $1.01\cdot10^9$ events if no truncation would be applied. The number of simulated undisturbed \ho events with $\SI{1.57}{\milli\second}\leq\dtch\leq\SI{10}{\milli\second}$ is $8.42\cdot 10^6$. Note that \pot is not considered in this simulation, as these events are already sorted out by the holdoff subfilter as discussed in \cref{sec:holdoff}.

    \subsubsection{Analysis of pileup-on-tail with both signals inside the time window}
    \label{sec:pot_analysis}
    A template fit as described in \cref{sec:templatefit} is applied to the generated traces, thereby obtaining the fit amplitude and \chitworeduced for each simulated event. The scatter plot of these fit parameters is shown in \cref{fig:amplitude_chi2}a. For $\chitworeduced\sim 1$, the line structure of the \ho spectrum is apparent with high abundances for fit amplitudes of $\sim 1.0$ (MI-line), $\sim 0.9$ (MII-line), $\sim 0.2$ (NI-line), $\sim 0.16$ (NII-line) and $\sim 0.025$ (O-lines). For larger values of \chitworeduced, arc-shaped structures that are centred around those amplitudes can be found.
    Three distinct groups of arc-shaped structures can be identified, culminating at $\chitworeduced \sim 10$, $\chitworeduced \sim 400$ and $\chitworeduced \sim 10000$. One finds that these groups correspond to \pit with amplitude $A_2$ corresponding to a \ho event from the O-lines, N-lines and M-lines respectively, while the shift of the arc-shaped structures along the x-axis depends on the initial amplitude $A_1$ of the pileup event.
    
    The structures can further be understood when considering the influence of the time difference between the pulses \dtch. For illustrative purposes, the path of decreasing \dtch for fixed amplitudes $A_1=A_2=1$ and relative sign of the polarities of $\Pi=+1$ is indicated in \cref{fig:amplitude_chi2}a. For $\dtch < \SI{1.57}{\milli\second}$, the value of \chitworeduced increases with decreasing \dtch and up to $\dtch\sim\SI{0.55}{\milli\second}$ the true amplitude of the triggered pulse is underestimated by an increasing amount. At $\dtch\sim\SI{0.55}{\milli\second}$, the largest value of \chitworeduced is reached. For further decreasing \dtch, the fit amplitude increases while \chitworeduced decreases up to the point where $\dtch=\SI{0}{\milli\second}$. Here, a fit amplitude of $A_1+\Pi A_2\simeq2.0$ and $\chitworeduced \sim 1$ is reached as it is expected for unresolved pileup of the two pulses. A corresponding mirrored structure arises from the same amplitudes with opposite relative sign $\Pi=-1$.
    
    \begin{figure}[!t]
         \centering
         \includegraphics[height=0.375\textwidth]{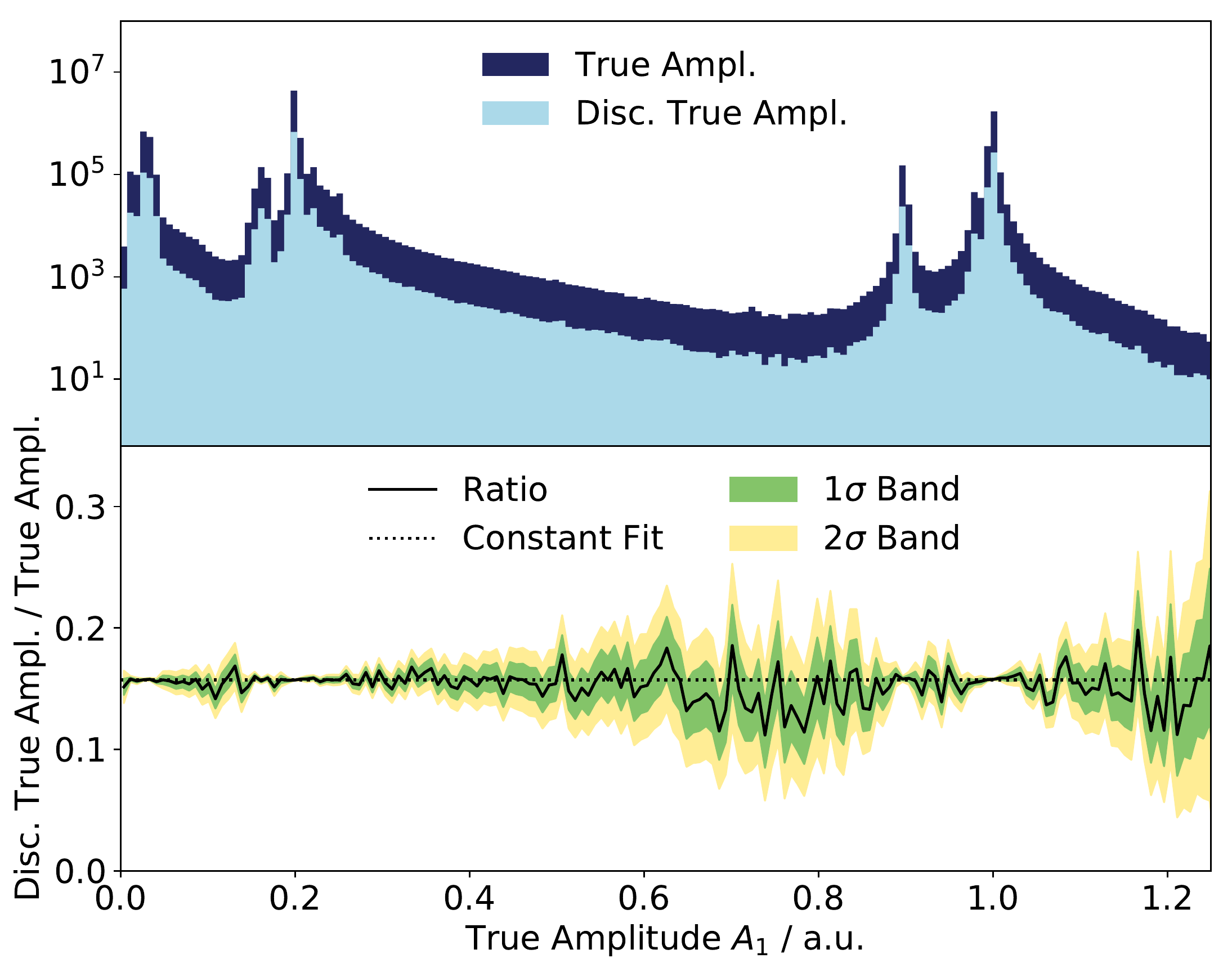}
         \caption{\textit{Upper panel:} Histogram of true amplitude $A_1$ of the simulated pileup events that are discarded by the \psfilter compared to the histogram of all simulated true amplitudes $A_1$. \textit{Lower panel:} Ratio of the two histograms and the corresponding uncertainty bands due to the Poisson error of the number of counts in each bin. The ratio agrees well with a constant fit, which indicates that \pit events are discarded by a \psfilter in an energy-independent way}
         \label{fig:energydependence}
    \end{figure}
     
    A simplified \psfilter that selects fitted traces with $\chitworeduced<1.3$ is used. Applying this filter to the simulated fit amplitudes yields a fairly clean theoretical \ho spectrum (see \cref{fig:amplitude_chi2}b upper panel orange), apart from a few outliers with fit amplitudes above 1.5 and below 0 that will be discussed in \cref{sec:energy_dep}. The region around the turning point in \cref{fig:amplitude_chi2}b at $\dtch\sim\SI{0.55}{\milli\second}$ is densely populated. This increase in density gives rise to a spiky structure in the histogram of the fit amplitudes of traces that are discarded by the \psfilter (see \cref{fig:amplitude_chi2}b lower panel). By comparing the scatter plot (\cref{fig:amplitude_chi2}a)with the histogram (\cref{fig:amplitude_chi2}b lower panel) one can associate the peaks at fit amplitudes of $\sim 0.9$ and $\sim 1.1$ to \pit of an MI-pulse ($A_1$) with an NI-pulse ($A_2$) for the two possible values of $\Pi$. In the same way, the peaks at fit amplitudes $\sim 0.7$ and $\sim 1.3$ correspond to \pit of two MI-pulses. Similar structures can be found centred around each line of the \ho spectrum.
    \begin{figure}[!t]
         \centering
         \includegraphics[height=0.375\textwidth]{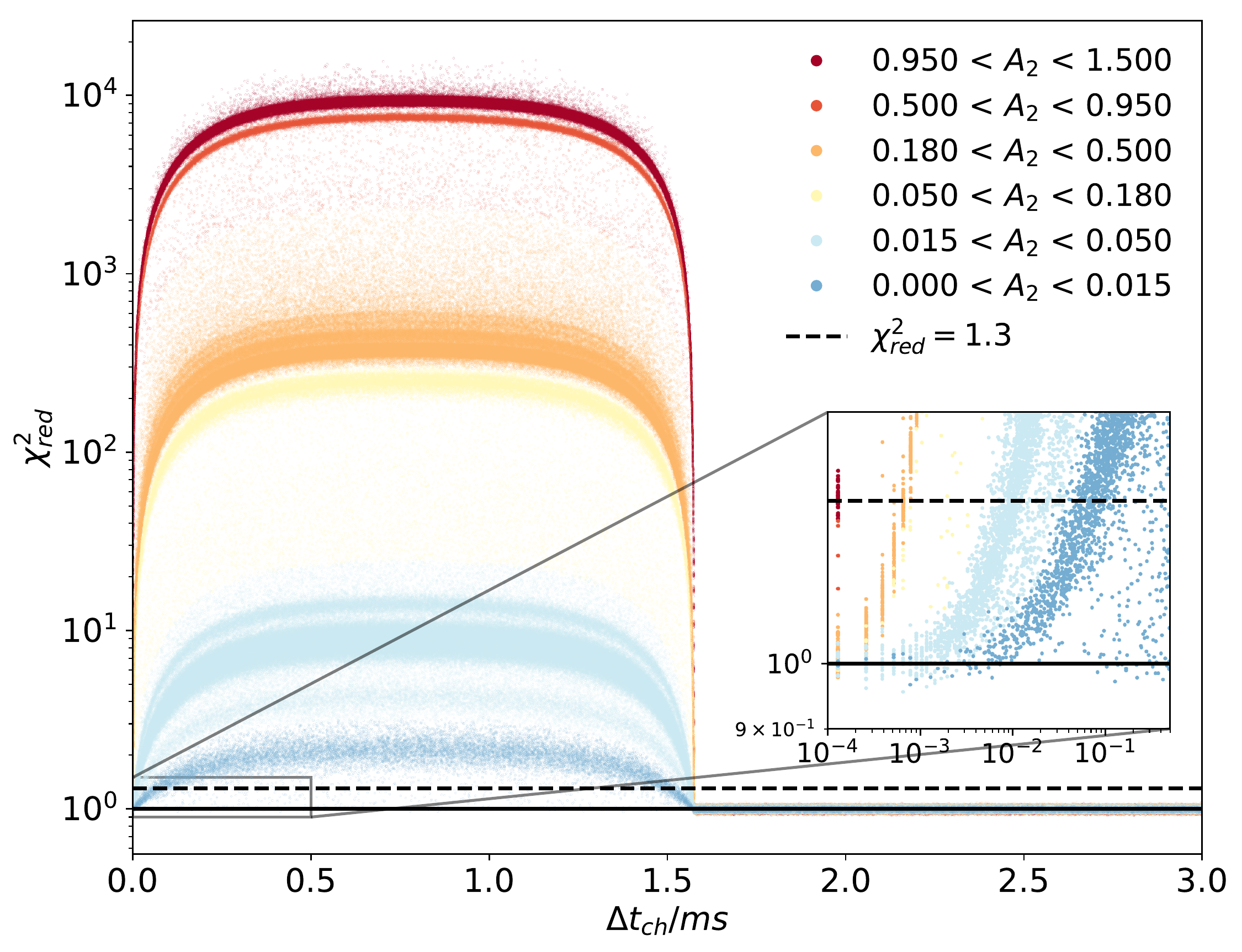}
         \caption{\chitworeduced as a function of \dtch for simulated \pit events. The value of \chitworeduced only depends on the amplitude of the pulse on the tail $A_2$ and \dtch but not on $A_1$. In the magnification it can be seen that it follows that the value of \dt for which \pit events satisfy $\chitworeduced<1.3$ depends on $A_2$. Note the logarithmic scale of the x-axis of the inset}
         \label{fig:dt_chi2}
     \end{figure}
    
    For the two-day dataset acquired with an ECHo-1k chip, the fit amplitude vs. \chitworeduced scatter plot of one detector channel (\cref{fig:amplitude_chi2}c) and the histogram of fit amplitudes discarded by a \psfilter for 18 detector channels (\cref{fig:amplitude_chi2}d lower panel) show structures that have striking similarities to the ones found in the simulated data. For better comparison, the same simplified \psfilter applied to the simulated data is also applied to the ECHo-1k dataset. The arc-shaped structures described above become apparent in the scatter plot for the data after applying the \tifilter. These in turn result in a similar structure of the histogram of fit amplitudes of traces discarded by the \psfilter. The most apparent difference between the histograms in \cref{fig:amplitude_chi2}b and \cref{fig:amplitude_chi2}d is the larger fraction of \pit events which arises from the truncated \dtch distribution used for the simulation. Furthermore, one can observe an asymmetry of spike pairs (e.g. fit amplitude of $0.9$ and $1.1$) in the histogram of the acquired data. For implanted detector channels, the activity in the two pixels is not identical, which yields to $P(\Pi = +1)\geq P(\Pi = -1)$. The probability that two consecutive triggers have the same polarity becomes larger for an increasing asymmetry of activity of the pixels. The asymmetry is maximal for detector channels, which only have one implanted pixel and thus $P(\Pi = +1) = 1$ and $P(\Pi = -1)=0$.

    \subsubsection[Energy dependence of a chi-squared-reduced filter]{Energy dependence of a \psfilter}
    \label{sec:energy_dep}
    
     \begin{figure}[!t]
         \centering
         \includegraphics[width=\linewidth]{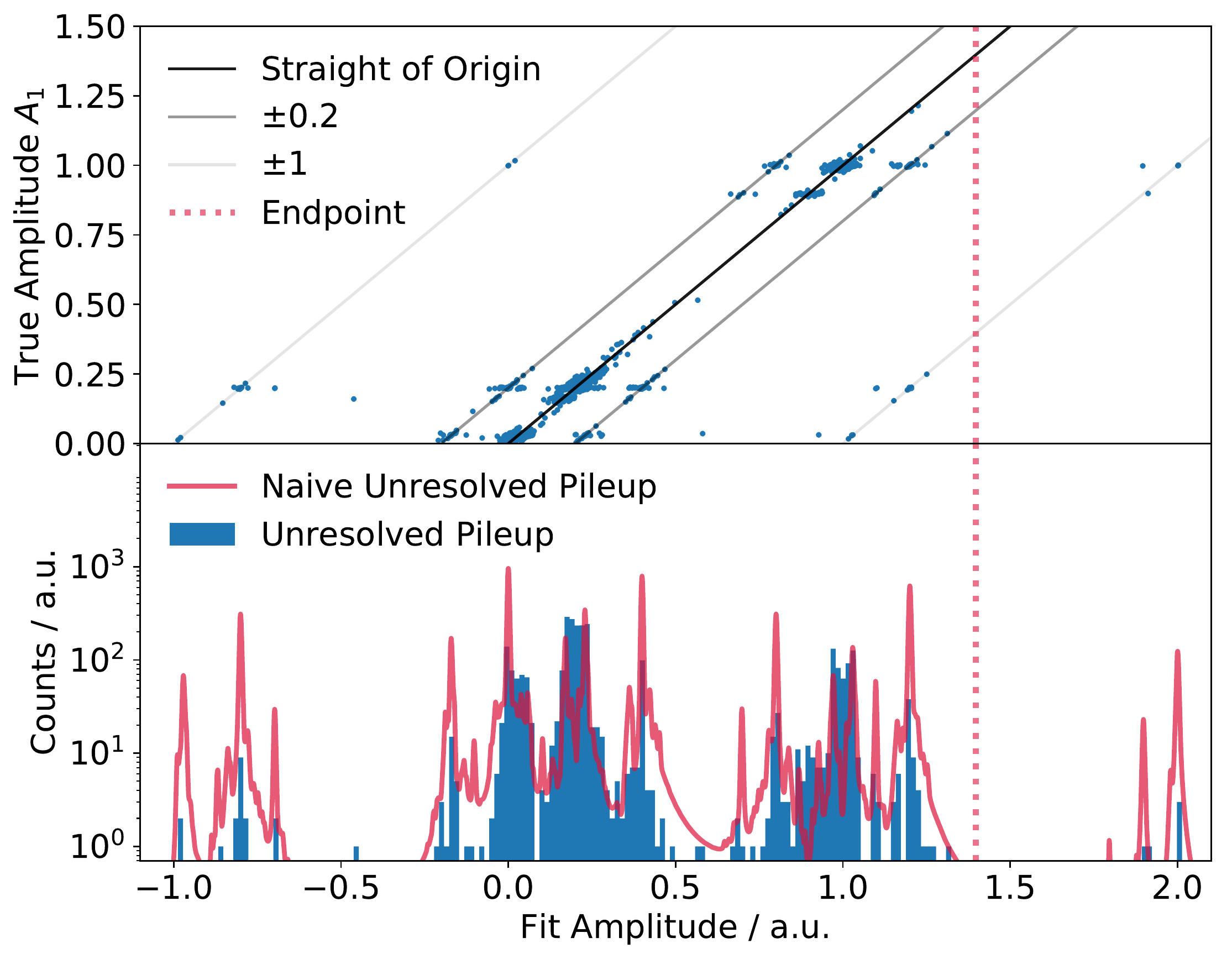}
         \caption{\textit{Upper panel:} Scatter plot of fit amplitude and true amplitude $A_1$ of each simulated unresolved pileup event. Most events are distributed around the straight of slope 1 through the origin. Additionally, straights of slope 1 shifted by $\pm 0.2$ and $\pm 1$ are drawn to guide the eye.
         \textit{Lower panel:} Histogram of the fit amplitudes of the events above. The autoconvolution of the theoretical \ho spectrum for a pileup fraction of $f_\mathrm{pu} =\SI{3e-6}{}$ is superimposed for comparison. In the simulated spectrum, structures from \pit with large amplitudes on the tail (e.g. at fit amplitude $\sim 2.0$) are more than an order of magnitude smaller than in the autoconvolution spectrum. In turn, unresolved pileup with barely altered fit amplitudes are more abundant in the simulated spectrum}
         \label{fig:unresolved_pu}
     \end{figure}
    
    In order to assess the energy dependence of a \psfilter, the histogram of true amplitudes $A_1$ of traces that are discarded by the filter is compared to the theoretical spectrum (\cref{fig:energydependence}). The ratio of the two histograms is shown in the lower panel together with the $1\sigma$ and $2\sigma$ error bands due to the Poisson error of the number of counts in each bin. A constant is fit to the ratio and, apart from one deviation of $-2.20 \sigma$ at an amplitude $A_1=1.06$, all ratios agree with the fit within the $2\sigma$ band. From this we can conclude that \pit events are discarded by a \psfilter in a fairly energy independent way.
     
    On a subdominant level, an energy-dependent distortion of the final spectrum arises from unresolved pileup, which in this context are pileup events that survive the \psfilter. In \cref{fig:dt_chi2}, \chitworeduced  is plotted as a function of \dtch. The data points are coloured according to the amplitude of the pulse on the tail $A_2$. Again, it becomes apparent that \chitworeduced only depends on $A_2$ and \dtch, but not on $A_1$. The horizontal bands indicated in \cref{fig:amplitude_chi2}a correspond to the location of the plateaus of \chitworeduced found for \dtch between $\sim \SI{0.25}{\milli\second}$ and $\sim \SI{1.25}{\milli\second}$.
    For smaller \dtch, the value of \chitworeduced steeply decreases towards $\chitworeduced=1$, as expected for $\dtch=\SI{0}{\milli\second}$. The inset in \cref{fig:dt_chi2} shows that the value of \dtch for which the events fulfil $\chitworeduced < 1.3$ is larger the smaller the amplitude $A_2$. These traces are considered good \ho traces by a \psfilter and thus correspond to unresolved pileup. For $0.015 < A_2 < 0.050$, i.e. OI-pulses on the tail, pileup is not recognised by the \psfilter for $\dtch \lesssim \SI{10}{\micro\second}$ while for $0.950 < A_2 < 1.500$, i.e. MI-pulses and higher on the tail, the time resolution for pileup is of the order of  $\dtch \sim \SI{100}{\nano\second}$\footnote{Note that the \dtch values given here depend on the \figureofmerit chosen for the simulated traces.}, which is of the same order as the time difference between two samples of a trace of \SI{128}{\nano\second}. This energy-dependent characteristic has the effect that the unresolved pileup spectrum does not simply correspond to the autoconvolution of the \ho spectrum as one would naively expect. Rather, we can infer from \cref{fig:dt_chi2} that the majority of unresolved pileup traces will feature small amplitudes $A_2$ and thus have a fit amplitude that deviates only slightly from their true amplitude. As a result, the acquired spectrum is only weakly distorted --- mainly by means of a slight broadening of the resonances. The spectrum of the reconstructed amplitudes of unresolved pileup events is shown in the lower panel of \cref{fig:unresolved_pu}. In the upper panel, a corresponding scatter plot of fit amplitude vs. true amplitude $A_1$  is presented. As expected, the majority of events are distributed near a straight line with unitary slope through the origin. These data points correspond to barely distorted traces from O-line pulses on the tail. Further accumulations can be found on the diagonals shifted by $\sim \pm 0.2$ (NI-pulse on the tail with $\Pi=+1$~(+) and $\Pi=-1$~(-)) and  $\sim \pm 1$ (MI-pulse on the tail). It can be seen that the outliers mentioned in \cref{sec:pot_analysis} with fit amplitudes above $1.5$ and below $0$ are concentrated near those shifted diagonals. The shape of the unresolved pileup spectrum is well understood and in particular no structure near the endpoint of the \ho spectrum emerges. A total of $2801$ unresolved pileup traces are found. The fraction of unresolved pileup for this simulation is $f_\mathrm{pu} = \SI{2.77e-6}{}$, considering that the number of simulated \pit events is equivalent to the number of \pit events from $1.01 \cdot 10^9$ events without truncating \dtch. For comparison, the autoconvolution of the theoretical \ho spectrum for a pileup fraction $f_\mathrm{pu} = \SI{3e-6}{}$ is superimposed in the lower panel of \cref{fig:unresolved_pu}. Unresolved pileup with an OI-line on the tail have similar rates in both spectra. However, structures with larger amplitudes on the tail are reduced by more than an order of magnitude, while those with barely altered fit amplitudes have a higher rate in the simulated spectrum.
    
    This simulation is representative for the estimation of unresolved pileup in the high statistics spectrum of ECHo-1k.

\section{Conclusions and Outlook}
In the ECHo-1k high statistics measurement, 58 MMC pixels, each loaded with an average of about \SI{0.5}{Bq} of \ho, have been operated over several months in order to acquire more than $10^8$ \ho events. This will allow to test the effective electron neutrino mass to a level of about \SI{20}{eV}. To reach this sensitivity, a new data reduction scheme has been developed. The aim of this scheme is to efficiently remove signals which could act as a background for the \ho spectrum, without sacrificing large fractions of \ho events and to precisely characterise any energy dependence of the filters.

We present a two-level data reduction scheme to obtain a clean signal from data acquired with ECHo-1k chips. The first level filter is purely based on the time information of traces. It is thus inherently energy independent. On a second level, the filtered data are further analysed by means of their deviation from a template pulse. The minor energy dependence due to unresolved pileup is well understood and can be modelled in an analysis of the \ho spectrum. All implemented algorithms are designed such that they can be applied online.

After the data has been filtered by the two-level data reduction scheme, the recovered amplitudes are corrected for temperature fluctuations of the entire setup. The energies of the events are then obtained by identifying the major resonances of the \ho spectrum and fitting their positions to the previously measured values with a polynomial function.

The methods discussed here can be adapted to be used for the next stages of the ECHo experiment. Future efforts will be directed towards resolving the energy of the first pulse in a \pit to maximise the signal yield of the second level filter. This is particularly important for a higher implanted activity per pixel, as envisaged in future phases of the ECHo experiment.
\medskip
\begin{acknowledgements}
We would like to warmly thank all member of the ECHo collaboration and the members of the low temperature group in Heidelberg for valuable and fruitful discussions. Special thanks to Josef Jochum and Alexander G{\"o}ggelmann.

The work described in this paper was supported by the DFG Research Unit ECHo under the contract ECHo GA 2219 / 2 - 2.
\end{acknowledgements}

\bibliographystyle{spphys}       
\bibliography{refs}
\end{document}